\documentclass[english,reqno,a4paper,12pt]{article}  

\pdfoutput=1

\usepackage[T2A]{fontenc}  
\usepackage[cp1251]{inputenc}  
\usepackage[russian]{babel}
\usepackage{graphicx,amsmath,amssymb,amsfonts,url}

\newtheorem{theorem}{\qquad Theorem}
\newtheorem{remark}{\qquad Remark}

\newcommand{\cB}{{\cal B}}

\newcommand{\cP}{{\cal P}}

\newcommand{\const}{\mathop{\rm const}\nolimits}

\begin{document}
\title {Synergetics and Its Application to Literature and Architecture}
\author{V.P.Maslov, T.V.Maslova}
\date{}
\maketitle
\selectlanguage{english}
\begin{abstract}
A series of phenomena pertaining to economics, quantum physics,
language, literary criticism, and especially architecture is
studied from the standpoint of synergetics (the study of
self-organizing complex systems). It turns out that a whole
series of concrete formulas describing these phenomena is
identical in these different situations. This is the case of
formulas relating to the Bose--Einstein distribution of particles
and the distribution of words from a frequency dictionary. This
also allows to apply a ``quantized'' from of the Zipf law to the
problem of the authorship of {\it Quiet Flows the Don} and to the
``blending in'' of new architectural structures in an existing
environment.
\end{abstract}
\section{Nonlinear addition and synergetics}

Let us begin with  the description of the Kolmogorov axioms for
nonlinear addition, which we will supplement by an additional
axiom.

A sequence of functions $M_n$ determines a {\it regular} type of
mean if the following conditions are satisfied (Kolmogorov).

\textbf{I.}  \  $M(x_1,x_2,\dots,x_n)$ is continuous and monotone
in each variable. To be definite, we assume that $M$ is monotone
increasing in each variable.

\textbf{II.} \  $M(x_1,x_2,\dots,x_n)$ is a symmetric function.

\textbf{III.} \ The mean of identical numbers is equal to their
common value: $M(x,x,\dots,x)=x$.

\textbf{IV.} \  A group of values can be replaced by their own
mean without changing the entire mean:
$$
M(x_1,\dots,x_m,y_1,\dots,y_n)=M_{n+m}(x,\dots,x,y_1,
\dots,y_n),\quad\text{where}\quad x=M(x_1,\dots,x_n).
$$

\begin{theorem}  \textrm{(Kolmogorov)}. If conditions {\rm I--IV} are
satisfied, then the mean $M(x_1,x_2,\dots,x_n)$ has the form
\begin{equation} \label{tag1}
M(x_1,x_2,\dots,x_n)=\psi\Big(\frac{\varphi(x_1)+\varphi(x_2)+
\dots+\varphi(x_n)}n\Big),
\end{equation}
where $\varphi$ is a continuous strictly monotone function, while
$\psi$ is its inverse.
\end{theorem}

For the proof, see \cite{3}.

It is rather obvious that any stable system must also satisfy the
following axiom, which is added to the Kolmogorov axiomatics
\cite{9}.

\textbf{V.} \  If to all the $x_k$ we add the same value $\omega$,
then the mean will increase by the same value $\omega$.

There exists a unique family of nonlinear functions satisfying
axiom~V. It has the form
\begin{equation} \label{tag2}
\varphi(x)=A\exp(\beta x)+B,
\end{equation}
where $A,\beta\ne0$, $B$ are numbers not depending on $x$.

Accordingly, the new addition of elements $a>0$ and $b>0$ is
defined by the formula
\begin{equation} \label{tag3}
a\oplus b=\frac1\beta\ln(e^{\beta a}+e^{\beta b}).
\end{equation}
\vskip-7pt\noindent
and the $\beta$-mean is\vskip-7pt\noindent
\begin{equation} \label{tag4}
\frac{a\oplus b}2=\frac1{2\beta}\ln(e^{\beta a}+e^{\beta b}).
\end{equation}
\vskip-3pt\noindent

Now we have obtained a family depending on the parameter $\beta$,
which for $\beta=0$ gives us the usual arithmetic, according to
which, in the biblical story described at the beginning of this
article, each brings the same amount of wine, while when
$\beta\to\infty$ everyone brings water:
$$
\lim_{\beta\to\infty}a\overset\beta\to\oplus b=\min(a,b).
$$
The arithmetic thus obtained led to the appearance of idempotent
analysis and tropical mathematics, now rapidly developing (see
\cite{10}, chap.~2; \cite{11}, sec.~9; \cite{12}, sec.~5.1).

The new arithmetic, which depends on the parameter $\beta$,
yields in the case of a large number of summands, formulas very
similar to those of thermodynamics. The parameter $\beta$ has the
physical meaning of $1/T$, where $T$ is the temperature. However,
the corresponding phase transition was not noticed by physicists.
In the works \cite{13, 14}, it was called {\it phase transition
of the zeroth kind.} Below we will discuss   this notion in more
detail.

\medskip

One of the fathers of synergetics, G.~Haken, in his article
\cite{1}, recalls the following story from the Ancient Testament:
``It was the custom in a certain community for the guests to
bring their own wine to weddings, and all the wines were mixed
before drinking. Then one guest thought that if all the other
guests would bring wine, he would not notice when drinking if he
brought water instead. Then the other guests did the same, and as
the result they all drank water.''

In this example, two situations are possible. In the first,
everyone contributes his share, giving his equal part, and
everyone will equally profit. In the second, each strives for the
most advantageous conditions for himself. And this can lead to the
kind of result mentioned in the story.

Two different arithmetics correspond to these two situations. One
arithmetic is the usual one, the one accepted in society, ensuring
``equal rights,'' and based on the principle ``the same for
everyone,'' for instance in the social utopia described by Owen.
In a more paradoxal form, this principle is expressed in
M.~Bulgakov's {\it Master and Margarita} by Sharikov: ``Grab
everything and divide it up.''

The aspiration to this arithmetic is quite natural for mankind,
but if society is numerous and non-homogeneous, then it can
hardly be ruled according to this principle. The ideology of
complete equality and equal rights, which unites people and
inspires to perform heroic deeds, can effectively work only in
extremal situations and for short periods of time. During these
periods such an organization of society can be very effective. An
example is our own country, which, after the destructions and
huge losses of World War II, rapidly became stronger than before
the war.

One of the authors personally witnessed such an atmosphere of
psychological unity when he was working on the construction of
the sarcophagus after the catastrophe of the Chernobyl nuclear
facility. The forces of the scientists involved were so strongly
polarized that the output of each of them was increased tenfold
as compared to that in normal times. During that period it was
not unusual for us to call each other in the middle of the night.

Nevertheless such heroism, self-denial, and altruism, when each
wants to give (and not to take) as much as possible, is an
extremal situation, a system that can function only for short
intervals of time. Here the psychological aspect is crucial,
everyone is possessed by the same idea --- to save whatever may
be saved at any cost. But the psychology of the masses, which was
studied by the outstanding Russian emigr\'e sociologist Pitirim
Sorokin, is presently studied only outside of Russia.

A similar enthusiasm takes hold of the masses during revolutions.
Thus, the October revolution in our country and the Great Chinese
revolution led the masses involved in the revolutionary process to
believe that they can ``move mountains.'' Nevertheless the policy
of military communism in Russia and the policy of the great leap
forward in China were doomed to failure.

In order for the system to survive, it was necessary to implement
a different ``egoistical'' arithmetic --- the NEP (New Economic
Policy) arithmetic. In China a similar arithmetic was in fact
introduced by Deng Siao Ping and his group.

This arithmetic is just as simple as ordinary school arithmetic
but, unfortunately, it is not studied in school.

The old arithmetic was linear, as the Soviet song goes ``We all
live happily, divide all things equally,'' whereas the new
arithmetic is that of nonlinear addition. And although it does
not seem objective, this arithmetic leads to certain rules which
ensure the establishment of equilibrium, the system becomes
self-organizing. Such an equilibrium exists, for instance, in
living nature. Everything follows the order of things (certain
animals eat other animals, a third group feeds on the first one
and so on), but, as we know, cataclysms also occur in nature.

The practice of corruption  has been established in Russian
society. It is a self-organizing  system. A person yields power
and enjoys wide opportunities. If the state were not to provide
him with benefits, he would obtain them from the people whom he
helps.

We have attempted to describe, on the basis of this new
arithmetic, both the equilibrium state of the system and the
catastrophic states that may occur. Equilibrium may last for a
long time, but progressively revolutions may arise. In order to
avoid catastrophic phenomena, during which the nonlinear
arithmetic explodes in a burst and is replaced by the ordinary
arithmetic, the political system of the state must be
sufficiently flexible.

As an example, let me mention the progressive implementation of
racial equality in the USA for Afro-Americans, which lasted for
approximately 30 years. If we look at this situation without bias,
then it is similar to the one described in I.~Erenburg's grotesque
novel ``DE Trust.'' Of course, clashes and even killings occurred,
but overall the system survived without serious cataclysms.

On the other hand, completely equalitarian systems, as we see from
historical examples, are usually extremely cruel to separate
individuals. Even the strongest and most natural feeling of love
can be severely repressed by such a system.

Let us recall the Russian rebel Stepan Razin, who was forced to
throw his beautiful wife overboard at the highest point of his
love. Stepan Razin was a rebel leader with colossal authority.
But the structure of which he was the leader demanded that he
perform this sacrifice and crime, and he complied. Another
example is the anarchist rule under the authority of the military
commander and agitator Nestor Makhno. When Makhno married his
great love (Vasetskaya) and their child was born, his cronies
arrested his wife and baby and forced them into exile under the
threat of death. Makhno suffered terribly, but did not leave his
comrades-in-arms.

One of the authors has also studied the very tenacious
anarcho-communist regime of Pol Pot and Chang Sari in Cambodia.
Under that regime marriages were impossible unless approved by
the communist cell of the bride-to-be. Following an assessment of
the qualities and faults of the potential wife, the bridegroom
was either allowed or forbidden to marry her, and another
prospective bride could be assigned.

In a similar way, during the war, military commanders could be
extremely cruel to young recruits experiencing the psychologically
very natural feeling of fear. In Cambodia, everyone ate at the
same table, and if someone, even a child, ate more than his
share, he would be severely punished. In the same way, in Russia
during the period of military communism, it was considered wrong
not to denounce anyone attempting to appropriate state-owned
resources (e.g., to steal a paper clip from one's office). In
contrast to this, at the present time, it is considered shameful
to inform the authorities, say, of a neighbor who sublets an
apartment without paying the corresponding city taxes.

Specialists in synergetics study problems related to revolutions
(see, for example, \cite{8}) and work on systems of preemptive
warning. Working on the economic problems set before him by the
industrialist I.Silaev (Deputy Chairman of the Council of
Ministers of the USSR, and later Chairman of the Council of
Ministers of the Russian Soviet Federative Socialist Republic),
one of the present authors and his specially recruited group of
collaborators developed such a system (see \cite{4, 5}). However,
this system was not supported by the subsequent government and
the group working on the problem was dissolved. As a result,
seven days before the 1991 putsch (which led to the
disintegration of the USSR), an article entitled ``How to avoid a
total catastrophe'' was published by the daily {\it Izvestia}.
The warning explained in the article was obtained via a
computerized mathematical~program.

All the collaborators were highly qualified experts and most of
them soon found jobs in foreign corporations and institutes.

\medskip

Now about history as a science. What is history? Suppose that a
person writes down the events of his life in a diary. Small
discrepancies from the uniform flow of life may be extremely
annoying for this person, but by the end of the day they may be
forgotten, small squabbles may then seem insignificant. If the
person compiles his diary only at the end of the week, and not
right after each event, his memory keeps only the important, the
essential ones, while to the lesser ones he no longer pays heed.
The same happens in the ``averaged'' economics. Prices in the
stock exchange jump very rapidly. These jumps depend on
unimportant events, on small political changes, but on the
average the economy evolves slowly, until a default, a crisis, or
a revolution occurs. Thermodynamical equilibrium also describes
the evolution of a system in time, but at small temperatures of
the system it ``waits'' until things ``level off,'' and only then
``records in its diary'' slightly changed state of the system.
However, abrupt phase transitions sometimes occur. It turns out
that there exist common laws governing all these
quasi-equilibrium states, namely the laws of synergetics.

Let us recall that a phase transition of the first kind for
physical processes is the transformation of matter from one state
to another (water to ice, water to vapor, etc.) during which the
volume and heat energy undergoes a jump, but there is no jump in
the ``kinetic'' energy. For example, water freezes at 0 degrees
Celsius and boils at 100. However, in the interval between 0 and
100 degrees, there exists a threshold temperature above which it
is impossible to save ice under any conditions. One can only
approach it. It is this threshold temperature that was called
temperature of the phase transition of the zeroth kind.

If this phase transition occurs, a large amount of ``kinetic''
energy is freed. For classical liquids, such a phase transition
has never been carried out in practice. It can only happen for
quantum liquids. From the mathematical viewpoint, phase
transitions of the zeroth kind in the thermodynamics of quantum
liquids, economic defaults, and revolutions in society are
phenomena of the same type. Namely, there is a limiting
``temperature'' of social tension such that, whenever it is
reached, there is no way that the revolutionary process can be
stopped; see \cite{5}.

Surprising as it may seem, {\it any} model of society, primitive
or complex, based on the $\beta$-arithmetic, leads, from the
mathematical point of view, to the same phase transition of the
zeroth kind, characterized by the freeing of kinetic energy and
branching of exactly the same type as in physics.

The fountain (burst) that accompanies this phase transition is
bloodletting, the ``fountain of blood'' produced by the French
revolution as it conquered huge territories. The hopes of the
left-wing bolsheviks for a world revolution were not realized
(their ``fifth column'' in the Antante was not strong enough) and
the kinetic energy (the burst) gushed only in the one huge
country.

But how must one choose the parameter $\beta$ (i.e., the
temperature), so as to achieve equilibrium, some sort of
egalitarianism, without carrying it too far?

It turns out that one can state a law called the lack of
preference law for a large number of ``distributed advantages,''
which will allow avoiding catastrophes for long enough. And we
shall see below that this law works in nature, in language, in
the organization of web sites, and in document systems, and there
is some hope that it may be implemented in our shaky society.

\section{The salary model}

Consider the following simple model. A personnel of $N$ persons
can receive the monthly salary of
$\lambda_1,\lambda_2,\dots,\lambda_n;\;\lambda_i>\lambda_{i-1}$.

The number of positions with salary $\lambda_i$ equals $Q/{i^s}$,
where $Q$ is sufficiently large, $s\ge0$.

The number of positions with given salary decreases as salary
increases.

Denote by $N_i$ the number of people receiving the salary
$\lambda_i$. The total expenditure on salaries is $\sum
N_i\lambda_i$. This number must not exceed the budget
$E$,\vskip-1pt\noindent

\begin{equation}\label{tag5}
\sum N_i\lambda_i\le E.
\end{equation}
\vskip-6pt\noindent

We will assume that all possible distributions of the personnel
among the $\sum_{i=1}^n Q/{i^s}$ positions is equiprobable.

What distribution $N_i$ is most probable and what is the
probability of deviation from this most probable number as
$N\to\infty,\,E\to\infty,\,n\to\infty$?

It turns out that the density of probability of the distribution
of $N_i$ satisfies the relation\vskip-1pt\noindent
\begin{equation}\label{tag6}
\frac{N_i}Q=i^s\frac1{e^{\beta(\lambda_i+\delta)}-1},
\end{equation}
\vskip-6pt\noindent
where $\beta$ and $\delta$ can be determined from the
relation\vskip-1pt\noindent
\begin{equation}\label{tag7}
\sum i^s\frac1{e^{\beta(\lambda_i+\delta)}-1}=N, \qquad \sum
i^s\frac{\lambda_i}{e^{\beta(\lambda_i+\delta)}-1}=E.
\end{equation}
\vskip-6pt\noindent
The relation \eqref{tag6} is similar to the
Bose--Einstein distribution law. Here $\beta$ is the same as the
one appearing in the definition of the new arithmetic. This will
be established below. Thus the parameter $\beta$ is determined
from a certain new equilibrium principle for all the situations
considered, the self-organization principle.

The papers \cite{6, 7} give the exact formulation of the
corresponding theorem, which encompasses the physical situations
as well as the ones coming from the humanities and the social
sciences.

First let us note the following physical phenomenon for ``weakly
nonideal Bose gas,'' namely, the Allen--Jones effect for liquid
helium-4. If we make a small hole in a narrow (of diameter
$10^{-4}$) tube through which liquid helium is flowing and raise
the temperature of that part of the tube to the critical level of
phase transition of zeroth kind, then a fountain (a {\it gerbe}
or burst) of height up to 14\,cm sprouts from the aperture near
the point of heating. This effect, discovered in 1938 by D.~Allen
and H.~Jones, is known as the fountain effect.

We shall assume that the real salary in the previous salary model
decreases linearly with the increase of the number of employees.
The coefficient of this decrease is $1/N$, hence any nonlinear
decrease in the leading term yields a linear one.

In this case, we obtain a mathematical picture coinciding with the
one appearing in the liquid helium-4 experiment, and for an
appropriate choice of the parameter $\beta$, a phase transition of
zeroth kind occurs.

Let us compute the mean salary using the $\beta$-mean. Denote
$$
q_i=Q/{i^s}.
$$
The number of different possibilities in which the number of
employees receiving the salary $\lambda_i$ was~$N_i$ can be
calculated by using the formula
\begin{eqnarray}
&& \gamma_i(k_i)=\frac{(k_i+g_i-1)!}{k_i!(g_i-1)!}\,;\label{tag8}\\
&&Q=\sum_{i=1}^ng_i,\label{tag9}\\
&&M(\beta,N)=\frac1{\beta
N}\ln\bigg(\frac{N!(G-1)!}{(N+G-1)!}\sum_{\{k_i=N\}}
\gamma(\{k\})\exp\left(\beta\cB(\{k\})\right)\bigg). \label{tag10}
\end{eqnarray}
\vskip-6pt\noindent
The mean salary is\vskip-6pt\noindent
\begin{equation}\label{tag11}
M(\beta,N)=\frac1{\beta N}\ln\biggl(\frac{N!(G-1)!}{(N+G-1)!}
\sum_{\{N_i\}_{\sum N_i=N}}\gamma_i(N_i)e^{\beta\lambda_i
N_i}\biggr).
\end{equation}

Invoking Stirling's formula and the Laplace method, we can
establish the connection of this mean as $N\to\infty$ with the
density of the distribution \eqref{tag6} \cite{16}.

In our example, we assumed that $q_i=i^{-s}Q$. Let us present the
argument in the general case.

\section{Dimension}

It is well known that the integration in spherical coordinates
over the space $\Bbb R^n$ leads to the density $r^{n-1}$, where
$r$ is the radius of the $k$-dimensional sphere. Similarly, as
shown by Yu.~Manin, to the fractional or negative dimension $D$
corresponds the density $r^{D-1}$.

The dimension of the space underlying a problem, as a rule, is
determined from the experiment or from some indirect
considerations. For example, the dimension of the point-like
presentation of a photograph (its digital version), like a
pointillist painting, is obviously of dimension zero, which yields
$s=D-1=-1$.

G.~Haken wrote that, in his opinion, an important role in
uncovering the causes of self-organizing systems must be played
by objects which simultaneously possess a quantum and a classical
structure. Such an object is, in particular, the quantum
oscillator. In it, the quasi-classical structure reduces to
finding an exact solution and performing a change of variables.
One can get rid of $h$, the Planck constant. Namely, for the
equation

\begin{equation}\label{tag12}
-\frac{h^2}{2m}\Psi^{''}_{xx}+\omega^2x^2\Psi=E_n\Psi,
\end{equation}
where the $E_n$ are eigenvalues, $\omega,m$ are constants
(frequency and mass), the substitution
$$
x=\big({\sqrt h}\big/{\root4\of{2m}}\,\big)y
$$\vskip-10pt\noindent
yields the equation\vskip-6pt\noindent
\begin{equation}\label{tag13}
-\Psi^{''}_{yy}+\omega^2y^2\Psi=\lambda_n\Psi, \quad
\lambda_n=\frac{2mE_n}h-\frac12.
\end{equation}\vskip-6pt\noindent

Let us put $\omega=1$; then $\lambda_n=n$.

Consider the line, the plane, and $3$-space. On the line, we pick
the points $i=0,1,2,\dots$, on the $x,y$-plane the points $(x,y)$,
where $x=i=0,1,2,\dots$; \ $y=j=0,1,2,\dots$ To this set of points
$(i,j)$ let us assign points on the line (natural numbers)
$l=1,2\dots$

To each point assign all the pairs of points $i$ and $j$ such that
$i+j=l$. The number of such pairs is $l+1$. On the $z$-axis let us
put $z=k=0,1,2,\dots$ so that $i+j+k=l$. Then the number $n_i$ of
points in 3-space will be 
$$
n_l=\frac{(l+1)(l+2)}2.
$$\vskip-6pt\noindent

In the first year of the 20th century Planck quantized the energy
of the oscillator, assuming that it changes by whole numbers
(multiplied by a parameter $h$, now known as the Planck constant).

If in formula ~\eqref{tag11}, we set $\lambda_i=i$, then in the
three-dimensional case to each $i$ will correspond a collection of
${i(i+1)}/2$ equal values $\lambda_i=i$ (they are the
multiplicities or ``degeneracies'' of the oscillator's spectrum).
The formula ~\eqref{tag7} in this particular case can be written
as\vskip-6pt\noindent

\begin{equation}\label{tag14}
\aligned N_l&=\text{const}\sum_{i=0}^l\frac{(i+1)(i+2)}{2(e^{\beta
i+\sigma}-1)}; \\
\Delta N_l&=\text{const}\frac{i(i+1)}{2(e^{\beta i+\sigma}-1)}
\quad\text{for large}\quad i, 
\\
dN&=\text{const}\frac{\omega^2d\omega}{e^{h\omega\beta}-1}
\endaligned
\end{equation}\vskip-6pt\noindent
(compare with formula (60.4) from
\cite{15}).

Here the weight $\omega^2$ coefficient is derived by Landau and
Lifshits from the fact that our space is three-dimensional.

It is more logical to argue differently. It is easy to verify, in
the $D$-dimensional case, that the sequence of weights
(multiplicities) of the number of variants $i=\sum_{k=1}^Dm_k$,
where $m_k$ are arbitrary natural numbers, have the
form\vskip-6pt\noindent
\begin{equation}\label{tag17}
q_i(D)=\const\frac{(i+D-2)!}{(i-1)!D!},
\end{equation}\vskip-6pt\noindent
and this constant depends on $D$.

For any $D$, we have\vskip-10pt\noindent
\begin{equation}\label{tag18}
N_l=\const\sum_{i=1}^l\frac{q_i(D)}{e^{\beta i}-1}.
\end{equation}\vskip-6pt\noindent
Comparison with experimental data shows that $D=3$, and so we may
conclude that photons ``live'' in three-dimensional space, and not
in four-dimensional Minkowski space, as one might have assumed.

Thus, for natural numbers, we have the sequence of weights $q_i$
(or simply the weight) of the form ~\eqref{tag17}.

It is known that some sets may have a non-integer dimension (some
attractors, fractals, the Sierpinski rug, etc.).

It is easy to continue the sequence of weights $q_i$ to the
general case by replacing the factorials by the
$\Gamma$-function:\vskip-6pt\noindent
\begin{equation}\label{tag19}
q_i(D)=\const\frac{\Gamma(D+i-1)}{\Gamma(i)\Gamma(D+1)}.
\end{equation}\vskip-6pt\noindent

Let us pick a negative $D$. This is the negative dimension of
space.

Suppose some person came into a huge inheritance $Q$ in different
forms and is wildly spending it left and right. If it is difficult
to estimate the amount of the inheritance, it is easy to assess
the amounts spent, which we assume increase with time $t$ as
$t^k$ (the appetite for spending may grow just like the appetite
for gain), then $k$ is the negative dimension, or $k$ is the
dimension of the ``hole'' that arises.

Moreover, here a new phenomenon, in the form of a condensate,
appears.

If $D>1$, then a condensation occurs in the spectrum of the
oscillator as $i\to\infty$, and a small perturbation suffices to
split the multiplicities, and the spectrum becomes denser and
denser with the growth of $i$. Negative values of $D$ and $D=0$
mean that as $i\to\infty$ the spectrum is noticeably rarified
(the constant in formula~\eqref{tag18} must be large enough).

For negative integer values of $D$, the terms $i=0,1,2,3,\dots,-D$
become infinite. This means that in the experiment they become
extremely large. This allows us to immediately determine the
negative (or zero) discrete dimension corresponding to the given
problem. There is a new condensate, which appears for small
normalized $\beta$.

In the work of Hagen and his followers, it is the
self-organization of {\it complex} systems which is considered.
Intuitively, the increase of ``complexity'' can be related to the
increase of entropy.

A.\,N.~Kolmogorov introduced the notion of complexity for discrete
systems, a notion which is finer than entropy, and showed that it
is close to the Shannon entropy.

The notion of the new arithmetic and the equiprobability of the
various variants that we consider here is simpler and more
precise than Kolmogorov complexity. In the simplest situations,
it not only coincides with Kolmogorov complexity, but also adds
certain specifics \cite{17}.

However, for the situation in which these notions considerably
modify probability theory \cite{6}, they give a lot more
information than Kolmogorov complexity.

As G.\,G.~Malinetskii and his collaborators correctly indicate in
their works (see, for example, \cite{18}), self-organization is
best observed in such empirical laws as the Zipf law and similar
rules. Many of these laws are related to natural numbers (the
number of people, the number of words) and this brings us to
linguistic statistics and semiotics.

\section{The Zipf Law}

First let us dwell on linguistic statistics.

Zipf empirically established the following remarkable dependence
between the occurrence frequency of a word in a corpus of texts
and its number in the ordering of words according to decreasing
frequency.

To frequencies correspond, in linear algebra, the eigenvalues of a
matrix. The dependence of the eigenvalues on their number is
rarely of simple form. As a rule, such a dependence is described
by very complicated relations.

In a frequency dictionary, to each word \footnote{The units of
frequency dictionaries may be different linguistic objects, for
simplicity we will call ``word'' any group (wordform or lexeme)
for which the dictionary indicates the occurrence frequency.}\
its occurrence frequency (i.e., the number of times that it
actually occurs in the texts \footnote{By occurrence frequency
linguists mean the absolute number of occurrences of the given
word in the corpus of texts under consideration, and not the
frequency with which it occurs (i.e., from the viewpoint of a
mathematician, the~number of appearances of the word divided by
the total number of words in the texts). } ) in the given corpus
of texts is assigned. Some different words may happen to have the
same occurrence frequency.

The analysis of frequency dictionaries shows that words appearing
in them can be split into three categories: (1)~``superfrequent''
(the so-called stop-words); (2)~frequently occurring words;
(3)~rarely occurring words.

It is surmised that words from each of the categories enjoy
unequal rights from the point of view of informativity, i.e., in
their reflection of the contents of the text. The first category
consists of words of the highest frequency. These are mostly
auxiliary words such as propositions and pronouns. They do not
play an essential role in clarifying the meaning of the text. If
one omits them, as in the text of a telegram, this usually does
not complicate the comprehension of the text. Many of them are
predictable, and therefore redundant. In computer science they
are called ``stop-words,'' and they are not taken into account.
As a rule, their distribution according to frequency (which turns
out to be quite chaotic) is not described by any algorithm which
could be used to indicate or specify their position.

One should, however, note that the words occurring in the super
high frequency zone are not considered ``useless'' from the point
of view of informativity by everyone. Thus, in the paper by
A.\,T.~Fomenko and T.\,G.~Fomenko \cite{19}, it is asserted that
the occurrence frequency of auxiliary words may be regarded as a
marker of the author's individual style. This conclusion was
obtained by the authors from empirical considerations in their
study of the frequency of auxiliary words in the works of twenty
Russian authors. They found that this frequency is stable within
corpora of texts of the same author and differs for different
authors. However, their study did not receive the recognition and
the support of the community of philologists.

One should also have in mind that in different languages (say
Russian and English), as well as in different versions of a
language (the language of grown-ups and children and teenagers,
in different types of slang), the role of auxiliary words in the
construction and the comprehension of texts may be quite
different. Thus in Russian (and other languages of flectional
type) the prepositions, which only repeat the meaning provided by
word endings, carry redundant information, whereas in English
(and other languages of analytical type), the role of
prepositions, adverbs, and other auxiliary words is rather
important (without them it is impossible to distinguish the
meaning of the so-called verb groups such as {{\it get in, get
out, get on, get off, get up}). On the other hand, in the
language of teenagers, an important role is played by, say,
``strengthening'' particles.

The second category consists of words with sufficiently high
occurrence frequency. Such frequencies are characterized by
lacunas, i.e., not all frequencies actually occur. Adapted
(simplified) texts usually consist of such words, because in such
texts rare words appearing in the original text are replaced by
more frequent synonyms or generic terms.

The third category of words of frequency dictionaries consists of
words of average frequency and rare words. In that part of the
dictionary, many words correspond to a given frequency, all the
frequencies are represented without lacunas.

The first (and main) Zipf law describes the second category of
words and is usually expressed in logarithmic coordinates
\vskip-6pt\noindent
\begin{equation}\label{tag20}
\ln n+\ln\omega_n=\const,
\end{equation}
\vskip-6pt\noindent
where $n$ is the rank of the word, i.e., its
number in decreasing order of frequency, $\omega_n$ is the
occurrence frequency of the word, i.e., the number of actual
occurrences of the word in the text. This formula means that the
product of the number (rank) of the word by its frequency is
(approximately) equal to a fixed constant.

Zipf himself noticed that in certain cases his formula is not
exact and that another constant, $D$ (which for frequency
dictionaries is close to 1), must be included:\vskip-1pt\noindent
\begin{equation}\label{tag21}
\frac1D\ln n+\ln\omega_n=\const.
\end{equation}
\vskip-6pt\noindent

The Zipf formula ~\eqref{tag20} in logarithmic coordinates
oversimplifies the actual relationship between frequency and rank.
Obviously the numbers $\log10^8=8$ and $\log(2\cdot10^8)=8+\log2$
differ very little, whereas the second of the numbers under the
logarithm sign, $2\cdot10^8$, is twice as large as the first one,
$10^8$. This means that the range of values of the right-hand side
in formula~\eqref{tag20} is considerably narrower than in the
formula without logarithms, as can be observed in the graphs
presented in~\cite{26}. In logarithmic coordinates the Zipf
formula is asymptotically correct, but is false in coordinates
without logarithms.

An attempt to give a linguistic interpretation of the constants
appearing in formulas ~\eqref{tag20} and ~\eqref{tag21} can be
found in the paper \cite{28}.

\section{Linguistic statistics.
A new viewpoint on frequencies} \vskip-4pt

In frequency probability theory, dual quantities appear: the
probability (or the number of occurrences of an event, a word, a
price, etc. divided by the total number of events, words, prices)
and values of the corresponding random variable. The sum of their
products is the mathematical expectation multiplied by the number
of trials, which, in the case of particles, is the total energy,
and in financial mathematics, the total capital.

Let $x_i$ be the values of the random variable, and $p_i$ be the
frequency probability of the occurrence of $x_i$. The number $N_i$
of particles at the level $\lambda_i$, divided by the total number
of particles, is the probability of ``hitting'' the level
$\lambda_i$.

If $N$ is the number of trials, then $N_i$ is the number of
occurrences of the value $\lambda_i$ of the random variable in a
sequence of $N$ trials, or the number of particles hitting the
energy level $\lambda_i$. If the number $N_i$ is large enough,
then the limit of ${N_i}/N$ is the probability of hitting the
energy ``level'' $\lambda_i$ in subsequent trials.

If we consider a game of dice, then $N_i$ is the number of times
the dice came up with the value $\lambda_i$ in $N$ throws or,
which is the same, in $N$ trials. In other words, $N_i$ is the
occurrence frequency or the frequency of hitting if we have not
fixed the total number of trials.

Thus, from the viewpoint of the proposed approach, having a given
frequency of occurrence and hitting the given energy level (for
particles) are essentially the same.

Now let us return to frequency dictionaries. The frequency of
``hitting'' is the occurrence frequency of a word in a group of
texts. Thus the occurrence frequency corresponds to the number of
particles (at a given energy level).

The number of words with the given frequency or with greater
frequencies is the value of a random variable. Indeed, if we
relate the informativity of words with their occurrence frequency
in the group of texts under consideration, the informativity
parameter will be applicable to words with given frequency or
higher. This is the usual assignment considered by linguists and
mathematicians working on frequency dictionaries.

However, here we are considering the opposite assignment, namely,
to the values of a random variable we assign frequencies, and to
probabilities we assign the number of corresponding words. Since
the sum of all these numbers of words is equal to the total
number of words (entries) in the dictionary, we can regard the
ratios of these numbers to $N$ as probabilities. In the
literature in linguistics, no such correspondence is indicated
(compare \cite{21}, p.~476; \cite{22}). From the linguistic point
of view, this correspondence appears to be overly formal.

Let us explain our point of view in more detail. We assume that if
$\omega_i$ is the number of occurrences of $k_i$ words
$\overbrace{a_1,\dots,a_{k_i}}$, then $\sum_{i=1}^n\omega_ik_i=M$
is the total number of words appearing in the texts on the basis
of which the frequency dictionary is constructed, while $\sum_{i = 1}^n
k_i=N$ is the number of words (entries) in the dictionary. Then
we can normalize the $k_i$'s, taking ${k_i}/N$ so that
$\sum_{i=1}^n{k_i}/N=1$, and then this can be regarded as the
probability of occurrence, while the number of occurrences
$\omega_i$ of the word $a_j$, $j=1,\dots,k_i$, is the value of the
random variable. At the same time, the value of
$\sum_1^n\omega_i$ is not known a priori. However, we may
calculate it for any dictionary. In fact, {\it a priori} we don't
know $n$ either, we must count it for each dictionary
additionally.

Therefore, the rather unusual point of view according to which the
number of occurrences is not a ``frequency,'' i.e., is a
non-normalized probability, is just as natural as the generally
accepted one.

We will present another formula, which gives a more precise
description of the relationship between frequency and rank. Since
the discrete character of the quantities under consideration is
essential for the derivation of the formula, we can say
that, in a certain sense, we are ``quantizing'' the Zipf law. In
other words, the formula derived below bears the same
relationship to the Zipf law as the formulas of quantum mechanics
relate to those of classical mechanics.

We will consider the frequency dictionary in ``inverse'' form: we
will number the words (entries) in increasing order of their
frequencies, beginning with the smallest one. Let us note two
important aspects. First of all, the frequencies are discrete. For
example, at the beginning of the ``inverse'' dictionary, each
successive frequency increases by 1. Secondly, the order in which
we list the words that have the same frequency is of no
importance: they can be listed in direct or inverse alphabetical
order or in any other order. It is only the number of words with
the given frequency which is important.

We stress that the proposed approach differs from the standard one
accepted in linguistics; it consists in the following. The
occurrence frequency of a given word (after normalization) is
standardly regarded as the probability of it occurring {\it in the
texts}. We look at the problem from the opposite point of view.
Suppose we have a frequency dictionary, which gives us fixed
frequencies. If we randomly choose a word (entry) {\it in the
dictionary}, what is the probability of ``stumbling'' on a given
frequency? If we choose the word from the alphabetical list of
entries, what is the probability of it having a given number of
occurrences? This probability equals the number of words that have
this occurrence frequency divided by the total number of entries
in the dictionary (the latter is denoted by $N$). Indeed, the
probability of having chosen a high frequency word, say, the word
``and,'' is extremely small, it equals $1/N$, whereas the
probability of randomly picking a word with occurrence frequency
1 will be the highest, because such entries have the highest
ratio in the list of entries.

In our considerations, frequency is regarded as the random
variable, while the number of words is the number of occurrences
of this random variable. Thus we will be using the ``inverse
order'' both for the entries of the frequency dictionary (ordered
by {\it increasing frequencies}) and in the relationship between
the random variable and the number of times it assumes a given
value. We speak of the probability of the occurrence in the
dictionary of a given occurrence frequency (i.e., the frequency
of occurrence of a given word in the collection of texts). We
call this probability {\it secondary}.

In the situation considered above, we chose words in the
alphabetical list of entries randomly and with equal probability.
However, the thesis that the choice should be equiprobable is
objected to by linguists. The objection boils down to the fact
that one mostly looks up unfamiliar words in the dictionary,
i.e., rare words. In that sense the vocables in a frequency
dictionary are informatively nonequivalent. Indeed, if we have a
corpus of texts, e.g., the collected works of a specific author,
on the basis of which some dictionary has been produced, be it an
alphabetical one, an encyclopedic one, a translation dictionary,
or a thesaurus, then can one assume that the user of this
dictionary, say, a translator, a literary critic, or just any
reader will choose words in it without any preferences? Clearly,
from the point of view of the reader, the vocables are not
equivalent: the more often a word occurs in the text, the more
familiar it is, the more understandable and thus the need to look
it up in order to obtain information about it is rarer. This is
the basis of the Ziva--Lempel code and the Kolmogorov theory as
developed by Gusein-Zade. And, conversely, the rarer are the
occurrences of the word in the text, the more probable it is that
one will need to look it up in the dictionary.

Therefore, when we consider a corpus of texts and the
corresponding dictionary, it is necessary to introduce a
preference function. Clearly, this function must be monotone with
respect to the occurrence frequency of words. Let us note,
however, that the extremely rare words must be excluded from our
considerations, so as not to concentrate too much attention to
them. In lexicographic practice one often introduces a lower
threshold on the occurrence of words. Thus, for instance, in the
compilation of dictionaries for information search systems
(information search thesauruses) terms with frequency below the
chosen threshold are not taken into consideration. In the
dictionary based on the British National Corpus \cite{20}
mentioned above, the only large frequency dictionary available
electronically, which we used in our study, only lexemes of
occurrence number not less than 800 and word forms of occurrence
number not less than 100 are presented. We will refer to the set
of words below the threshold as the {\it condensate}.

We are considering the frequency probability theory and we impose
an additional condition on the random variables. Namely, we
assume that the set of values of the random variable (i.e., real
numbers $x_1,\dots,x_s$) is {\it{ordered}}. Usually, in
mathematical statistics, one establishes an order in the values
of a random variable in accordance to their size (``order
statistics''). Among the numbers $x_1,\dots,x_s$ there may be
equal ones (\cite{23}, p.~115). In that case, one usually
amalgamates them and takes the sum of the corresponding
``probabilities,'' i.e., the ratio of the number of ``hits'' of
$x_i$ to the total number of trials. But if we introduce a common
order relation, we can no longer amalgamate equal values of
$x_i$. To a certain extent we can say that the family
$x_1,\dots,x_s$ is a finite ``time series'' ({\it loc. cit.}
p.117).

The difference between this probability theory and the classical
one is also due to the fact that not only the number $N$ of
trials can tend to infinity, but also $s$. Moreover, if the time
series is infinite, then $s=\infty$. Let $n_i$ be the number of
``occurrences'' of the value $x_i$, then\vskip-1pt\noindent

\begin{equation}\label{tag22}
\sum^s_{i=1}\frac{n_i}Nx_i=M,
\end{equation}
\vskip-6pt\noindent where $M$ is the mathematical expectation.
Note that the left-hand side of formula \eqref{tag22} contains the
``scalar product'' of pairs, one of which is normalized, i.e.,
$\sum^s_{i=1}{n_i}/N=1$, or, as physicists say, the family
$\{n_i\}$ is a ``specific quantity.'' In thermodynamics, it is
known as an extensive quality, while its ``dual'' is an intensive
one (e.g., volume--pressure).

The cumulative probability $\cP_k$ is the sum of the first $k$
probabilities in the sequence $x_i$:\vskip-1pt\noindent
$$
\cP_k=\frac1N\sum_{i=1}^kn_i, \quad \text{where} \quad k<s.
$$\vskip-6pt\noindent

Essentially, the cumulative probability is the ``distribution.''
In other words, distribution necessarily requires the ordering of
the values of the random variable or the notion of finite time
series.

\section{Linguistic statistics. The logarithmic law}

Since the number of words occurring only once constitutes no less
than $1/3$ of the dictionary, we can assume that the dimension is
zero, i.e., in the formulas we can set $D=0$. This leads us to an
expression of the form \cite{7}:\vskip-10pt\noindent

\begin{equation}\label{tag23}
N_l\cong T\ln\frac{\omega_l}{1+\alpha\omega_l},\ \
T,\alpha=\const.
\end{equation}
Its derivation as $\beta\to0,\sigma=0$ is
trivial:\vskip-6pt\noindent
$$
N_l=\frac1\beta\sum_{i=1}^l\frac1{\omega_i+\alpha
\omega_i^2}+O\Big(\frac1{\beta^2}\Big).
$$
Since $l>\varepsilon N,$ $\Delta\omega_i=1$, it follows that, as
$N\to\infty$,\vskip-1pt\noindent
$$
\sum_{i=1}^l\frac1{\omega_i+\alpha\omega_i^2}
=\int^\omega\frac{d\omega}{\omega+\alpha\omega^2}=
\ln\frac\omega{1+\alpha\omega}+c,
$$\vskip-8pt\noindent
and therefore\vskip-5pt\noindent
$$
N_l=T\ln\varrho\frac{\omega_l}{1+\alpha\omega_l},\qquad T=\frac
1\beta, \quad\text{where}\quad\varrho=\exp(c/T).
$$\vskip-4pt

Now let us pass from the numbers of rarely occurring words to the
most frequently occurring words among the $N$ words involved in
our formula. This leads to what linguists call the ``rank'' of a
word.
Clearly, the rank is the number $r_l=N-N_l$. Hence
$$
r_l=T\ln(1+1/({\alpha\omega_l})).
$$\vskip-3pt\noindent
If we pass to $r^0=T,\omega^0=1/\alpha$, we finally obtain
$$
{r_l}/{r^0}=\ln(1+{\omega^0}/{\omega_l}).
$$\vskip-3pt\noindent
Here $r^0$ and $\omega^0$ may be regarded as a normalization. We
have obtained the {\it logarithmic law\/}:

{{\it one can normalize the rank and the occurrence frequency in
such a way that the normalized rank $r$ and frequency $\omega$
will satisfy the relation}}\vskip-1pt\noindent

\begin{equation}\label{law}
r=\ln(1+1/\omega),
\end{equation}
\vskip-3pt\noindent {{\it where $r$ is the normalized rank of the
word while $\omega$ is its normalized occurrence frequency.}}

For $\omega\to\infty$, \ $r\sim1/\omega$, this is the Zipf law.
The corresponding plots appear in the articles \cite{24,
arX_Neg_Dim, arX_Hole_Dim}.

\begin{remark}
The Zipf law is usually written in double logarithmic
coordinates. It expresses the value of the logarithm of the rank
of a word in the frequency dictionary corresponding to the given
corpus of texts $M$ (recall that the rank is the number of the
word in the list of all words arranged in order of decreasing
frequency) in terms of the logarithm of the occurrence frequency
of this word in $M$. This relationship is shown by the hypotenuse
of a triangle with right angle formed by the coordinate axes.
\end{remark}

This fact is reminiscent of the dequantization procedure (see
\cite{SecondDequant}.).

It was shown in \cite{5} that the Zipf law corresponds to the
dequantization of the Bose--Einstein distribution law in
dimension zero.

Under the analytical continuation to zero of the dimension, a pole
appears \cite{SecondDequant}, and this pole can be interpreted as
a condensate of words. Actually, for books, the number of words
occurring exactly once is no less than a third of the total
number of words in the corresponding dictionary. From this, we
can conclude that the condensate is just the one pole, and this
implies that the dimension of each oscillator of the system of
$N$ oscillators is equal to zero.

This, in its turn, means that the straight line of the plot in
double logarithmic coordinates forms 45 degree angles with the
coordinate axes.

We begin ordering the words in formula \eqref{law} beginning with
rank 2, i.e., above the condensate. If we return to the ordering
proposed by Zipf, i.e., counting words beginning with the highest
frequency $R$ (i.e., $R=N-r$, where $N$ is the total number of
words, except those in the condensate, appearing in the
dictionary), then, from \eqref{law} we will obtain a relationship
that is better expressed in terms of $w(R)$ (rather than $r(w)$).
This relationship has the form
$$
w=\frac{\const}{e^{\beta'R+\beta'\chi'}-1},
$$\vskip-4pt\noindent
and as $\beta\to0$ becomes the Zipf law.

The constants $\beta',\,\chi'$ are expressed via the chemical
potential and the temperature.

Now the question of the length of ``adapted'' (simplified) texts
arises. Suppose we have thrown out all words of occurrence
frequency less than $w_0$. What will the volume $V(w_0)$ of the
remaining text be? The corresponding formula, which follows from
the Bose--Einstein distribution in dimension zero, can be written
as\vskip-1pt\noindent
$$
V(w_0)=\frac1\beta\int_{w_0\beta}^\infty\,\frac1{e^{x+\beta\chi}-1}\,dx,
\quad \beta\ll1,
$$\vskip-4pt\noindent
where $\beta$ is the inverse temperature, $\chi$ is the chemical
potential.

This formula gives the exact answer up to
$o(1/\sqrt{\beta|\ln\beta|})$.

\begin{remark}
If the dimension is equal to one, as in the
case of the occurrence frequency of ``Japanese candles'' on the
stock market (see \cite{5}), then
$$
R(w_0)=\frac1\beta\int_{w_0\beta}^\infty\,\frac1{e^{x+\beta\chi}-1}\,dx.
$$
\end{remark}

\section{Linguistic statistics. Attribution of texts}

Statistical properties of the structure of natural language texts
have long been in the center of interest of linguists as well as
mathematicians. Numerous studies in linguistic statistics have,
in particular, applications to literary criticism and yield
certain tools used in the attribution of texts, i.e., in
determining the author of a given text (see \cite{25} and the
references therein).

Many studies have been devoted to the attribution of the text of
M.~Sholokhov's novel {\it Quiet Flows the Don}. In the present
paper, we present a new approach to this question, based on some
of our recent work \cite{5, 26, 28, 24}.

Everyone knows what a enciphered text or an encoded telegram is.
Cryptography studies the decoding of enciphered texts and teaches
how to encode texts so as to make it difficult to decipher them.
How is this related to literature?

Abstruse language, the coining of new words, sound-related
associations are in fact encodings of what the poet or writer
wishes to ``secretly'' tell the reader: feelings and associations
that must appear in the reader's mind as the result of a
combination of sounds, i.e., by means of {\it onomatopoeias} or
{\it alliterations}. Here is an example:

{\it Toads nest there and small snakes hiss} (K.~Balmont, {\it In
the Reeds}).

Why does the reader feel that he is touching something cold and
slimy? This feeling is not explicit in the text, it is ``encoded''
in the combination of sounds.

In the above example, the language is unbroken. But one can break
up sentences and words into pieces as V.~Khlebnikov does, and this
may have an even stronger effect. In such intricate verse, in such
abstruse language, the poet enciphers a device that arouses
specific feelings in the reader's consciousness. Possibly,
Khlebnikov's esoteric and bewildering play ``Zangesi'' should be
staged by a producer and played by actors who are capable of
entirely deciphering Khlebnikov's abstruse language.

It is clear that the science of cryptography and the deciphering
of literary texts by critics are related, and the approaches to
this subject matter are clear.

But much more complicated problems confront the mathematician in
studying the science called ``stegonography.''

It is one thing when we are given an encrypted text that we must
decipher, and a completely different one when we study an object
which possibly contains some secret data, for instance, we are
studying a short story, a painting, some statistical data that
possibly contain encoded secret information, incorporated in such
a way that no one has guessed that it is there.

Speaking of Leonid Andreev, Leo Tolstoy wrote: ``He tries to scare
me, but I'm not frightened.'' Sometimes you have the opposite
situation --- the writer does not try to scare you, but you are
frightened. Where has he hidden the secret writing that frightens
the reader, although there is nothing frightening in the text?
This is a question that literary critics find hard to answer. But
sometimes it is possible, by means of a thorough analysis, to find
the hidden strings that the artist used, consciously or
unconsciously, to create his work. And when it is impossible to
decipher these hidden strings, while the effect is so strong that
it seems divine, then we say that we are faced with the creation
of a genius. Thus the mystery of Mona Lisa (Joconda), the mystery
of the femininity of Giorgione's Venus, the mystery of Nike tying
her sandals, strikes our imagination so much that we want to cry
out: ``it cannot be.'' It cannot be that this was created by a
human being.

It is even more difficult for contemporaries to realize that this
was created by an acquaintance, someone with whom one contacts in
a familiar atmosphere. Colleagues may even acquire the ``Salieri
syndrome.''

Note also that Da Vinci and Giorgione have no other similar
paintings, the ones mentioned stand out even in the work of these
two great painters. Where does the effect produced by these
paintings disappear when we look at their reproductions? Those who
had seen the originals with their own eyes and then admire their
copies, feel the effect produced by the original for a while, but
then after a while the magic disappears entirely.

And the translation of a work of literature to another language,
isn't it the analog of the copy of a painting? How much of the
charm is lost in translation?

So how can we decipher this ``secret writing'' with its mysterious
effect? Can the methods of steganography and, more generally,
those of the mathematical approach help?

Craftsmanship can be perfected and developed, it can be explained.
Inspired vision, creative illumination strikes suddenly and no
one knows from where. N.~Gumilev perfected his craft and taught it
to others, but Alexander Blok was struck by a sudden illumination
and wrote his great poem {\it The Twelve} in one night, at one
sitting. This poem has no similarity whatever with the rest of
Blok's poetry. How can that be explained?

Sholokhov's {\it Quiet Flows the Don} is the same kind of
mysterious work of literature. The enigma of that novel is in
that its author succeeds in fitting life with all its nuances
into two volumes of prose so that the reader feels that has seen
all of it with his own eyes, lived in the same village and
breathed the same air as Grigory Melikhov, was involved in the
complex relationships between the villagers, experienced the
euphoria of successful revolt when life became unbearable. And,
most important, the reader feels that he has watched all the
events, seen all the details with his own eyes, has apparently
missed nothing. Despite the tragic ending, the novel leaves a
bright feeling of purification, one wants to plunge into that
life and that epoch once more, rereading the book again and again.

It is hard to believe that a human being created such a
cinema-novel, every shot of which is not a photograph, but a
three-dimensional picture by an outstanding master; created in
such a way that the reader experiences each scene with all five
senses, including smell and touch. To try to understand these
hidden strings, let us look at one of the scenes of {\it Quiet
Flows the Don}, that of the wedding of Grigory and Natalia. It is
written up with the rich brush strokes of the Dutch masters,
colorfully, naturalistically, delectably. The reader is perhaps
satiated, as the other guests, by the rich food offered at the
wedding. The description has barely noticeable strings that lead
to the sudden flashes occuring during the drunken stupor of the
wedding fiest, flashes that can only occur in real life. When
does this happen? When, in that drunken stupor, two-three
familiar and well-loved faces fly past. Remove these flashing
faces, and you are left with just a lovely picture of a wedding,
something for tourists to admire, but the feeling of being there,
of belonging, of closeness, of something in one's soul disappear.

M.~Bulgakov, in a review of writings by Yuri Slezkin, wrote that a
writer should love his characters.

Sholokhov loves Grigory, loves Petro, loves Dunyasha, loves
Natalia, and loves\dots the Don river. During the wedding, at
some moment Dunyasha's eyes sparkle, Prokofich glances at the
Don, Grishka secretly makes a face, while Petro chuckles amiably
at his younger brother. Throw out these little episodes, omit
these little pieces of the text, and all the things that sparkled
and shone in changing colors like droplets of dew in the morning
sun, giving volume to the narrative and placing the reader into
that volume, and all these things will disappear.

It is impossible to explain how Sholokhov achieves this effect.
But if the ephemeral bits of text mentioned above are left out,
the effect disappears. This is similar to the mystery of Mona
Lisa. It is these little bits that contain the mystery which
distinguishes the original from the copy. We know that these
little bits contain the secret writing, but we are unable to
decipher its nature. But the fact that we have identified where
the secret writing appears, means a great deal from the point of
view of steganography.

Yes, Sholokhov loves his Grigory, although he shows that the
latter is in the wrong and blames him for that. Because of his
love of a hero opposed to the bolsheviks, Soviet critics branded
the novel antisoviet.

No single writer could produce a second epos of the same scope ---
just as it is impossible to live two lives.

Soon after the publication of the first part of the book, a legend
appeared, claiming that Sholokhov had found the manuscript on the
body of a dead White Army officer and plagiarized it. The supposed
``real author'' was even named by some --- it was the well-known
(at the time) writer Kryukov, author of the novel {\it Zyb'}
(``Ripples''). After the appearance of the second part of {\it
Quiet Flows the Don}, this version was automatically abandoned.

However, in 1974 this version surfaced again. It was supported by
authoritative public figures, including former dissidents.

Some of them had seen, in political forced labor camps, how
manuscripts were carefully hidden and how the authors, aspiring
to save the manuscripts after their death, passed the text to
fellow prisoners. If the manuscript got into the hands of an
informer or an officer of the secret police, the latter would
keep it and even show to others, thus provoking the expression of
an opinion. Apparently, for this reason such behavior was deemed
natural outside the camps as well. But plagiarism of this scope
never occurred in the history of literature. It is only jokingly
that the Goncourt brothers explained how they created their work:
one brother would run from one publisher to the other, while the
other one would sit at home guarding the manuscript so that it
would not be stolen.

It did happen that literary ``slaves'' were hired to help the
writer (Alexandre Dumas did this in France, Surov, in the USSR,
for which he was thrown out of the Writer's Union), some authors
edited or modified the work of others, as did A.\,N.\,Tolstoy in
transforming the Italian children's book character Pinnochio into
the Russian Buratino, or as the once demagogical popular writer
Misha Evstigneev rewrote Gogol's work ``for the masses.''

But these examples were the rewriting of published works, not
unpublished manuscripts. Stealing manuscripts can only happen in
the mind of science fiction writer or a person prone to fantasy
and believing in his own fantasies.

Now the question of attributing a text to someone, when we are
speaking of events long past, is a different story. Here we are
nor dealing with plagiarism or theft.

But {\it Quiet Flows the Don} is not a short story or a small poem
born in a political labor camp, it is a huge epos that took
Sholokhov 15 years to write. How can one imagine that anyone could
have found a manuscript and would progressively present it, year
after year, as his own work?! F.\,F.~Kuznetsov succeeded in
proving that Grigory had a real life prototype; he also showed
how {\it Quiet Flows the Don} was actually written. Finally,
Sholokhov's drafts of the novel were recently discovered, and it
would seem that the question is closed. However, Sholokhov's
opponents (those who support the plagiarism version) have not
expressed their views on that discovery, have not publicly
admitted that they were severely mistaken concerning the great
Russian writer. Thus the situation remains unclear, especially
for those who respect the authority of Sholokhov's opponents and
believe whatever they say, irrespective of any logic.

Sholokhov's second novel, {\it Raising the Virgin Lands} is
indeed a very different book, written as by an impartial
observer, not ``from the inside,'' but ``from the outside.'' The
lands surrounding the basin of the river Khopr (a confluent of
the Don) was populated by a historically younger generation of
Kossaks, without the same deep roots. During the Civil War, they
were on the side of the Red Army.

Besides, Sholokhov was carrying out a political mission --- to
write a novel about the role of the communist party in the
villages. Sholokhov used to go hunting in these places, he knew
the local people, but was not part of them as he was part of the
Kossacs from the villages around Veshinskaya (where the action of
the first novel takes place). Nevertheless, we will see below,
that from the point of view of scientific text analysis, there is
much in common in the two novels.

In {\it Raising the Virgin Lands}, there is no hero whom Sholokhov
loves with all his heart, there is no such person, but there is
nature, the landscape of the Don, without its most important part,
the Don itself, but then nature appears in all of Sholokhov's
writings. So we have chosen, for our comparison, the descriptions
of nature in his two novels ({\it cf.} the work of
F.\,F.~Kuznetsov about nature in Sholokhov's fiction).

We have performed a computer analysis, first of all, of
Sholokhov's descriptions of nature, then, after deleting the
parts of {\it Quiet Flows the Don} dealing with the author's
favorite characters, we have compared this truncated version of
his first novel with {\it Raising the Virgin Lands}. One can see
with the naked eye, without any mathematical analysis, that these
texts were written by the same hand.

For the mathematical comparison, it turns out that one must use
the part of the plot near the place where the curve bends. The
corresponding words are not auxiliary words, nor are they very
rare words, they are words which occur often enough in most parts
of the text.

Let us compare the constants in the logarithmic law with those in
the Bose--Einstein distribution:

\begin{equation}\label{tag24}
rang=\int_1^\omega\frac{dx}x
\Big(\frac1{e^{\beta(x-\mu)}-1}\Big)+\alpha.
\end{equation}
Here $\alpha$ is the correction to the rank, $\beta=1/\Theta$ is
the inverse temperature, $\mu$ is the chemical potential divided
the temperature, so that we have three independent parameters. For
$\beta\omega\ll1$ in this interval, we have
\begin{equation}\label{tag25}
rang=\int_1^\omega\frac1x\Big(\frac1{(1+\beta x)
e^{-\beta\mu}-1}\Big)dx+\alpha.
\end{equation}

Denoting $e^{-\beta\mu}=d$, let us compute the obtained integral
\begin{equation}\label{tag26}
rang=\frac1{d-1}\log\biggl(\frac{(1+\frac{d\beta}{d-1})\omega
g}{\frac{d\beta}{d-1}\omega+1}\biggr),
\end{equation}
where $g=(d-1)e^\alpha$. Comparing with
\begin{equation}\label{tag27}
rang=a\log\frac{b\omega}{c\omega+1},
\end{equation}
where $a,b,c$ are determined by experiment, we obtain

$$
a=\frac1{d-1},\qquad b=\Big(1+\frac{d\beta}{d-1}\Big)g,\qquad
c=\frac{d\beta}{d-1}.
$$
We find

$$
d=\frac1a+1,\qquad \beta=\frac{c(d-1)}d,\qquad g=\frac
b{(1+\frac{d\beta}{d-1})}.
$$
This allows us to express the parameters of the initial
distribution.

$$
\beta=\frac c{a+1},\qquad
\mu=-\frac{a+1}c\log\Big(1+\frac1a\Big),\qquad
\alpha=\log\frac{ab}{1+c}.
$$

We obtain the following values of the parameters:
\begin{align*}
&\text{{\it Quiet Flows the Don} (different parts)}\\
&\hskip4cm\text{ -- inverse temperature $\beta$: }&\quad& 2.667,
\,\, 1.684, \,\,1.956, \,\,1.684, \,\,1.683
\\
&\text{{\it Raising the Virgin Lands} \ \,-- inverse temperature
$\beta$: }&\quad& 1.906
\\
&\text{{\it Zyb'} (Kryukov's novel) \quad \ \,-- inverse
temperature $\beta$: }&\quad& 0.604
\\
&\text{{\it Quiet Flows the Don} (different parts)}\\
&\hskip3.95cm\text{ -- chemical potential $\mu$: }&\quad
-&0.0173, \,\,-0.0053, \,\,-0.0036, \,\,-0.0053
\\
&\text{{\it Raising the Virgin Lands} \ -- chemical potential
$\mu$: }&\quad -&0.0065
\\
&\text{{\it Zyb'} \hskip3.3cm -- chemical potential $\mu$: }&\quad
-&1.519
\\ 
&\text{{\it Quiet Flows the Don} (different parts)}\\
&\hskip3.95cm\text{ -- addition to the rank $\alpha$: }&\quad&
0.6414, \,\, 0.4216, \,\,0.4956, \,\,0.4216
\\
&\text{{\it Raising the Virgin Lands} \ -- addition to the rank
$\alpha$: }&\quad& 0.4848
\\
&\text{{\it Zyb'} \hskip3.3cm -- addition to the rank $\alpha$:
}&\quad& 0.5435
\end{align*}

Thus we see that the chemical potential of {\it Zyb'} is hundred
of times more than that of {\it Quiet Flows the Don} and of {\it
Raising the Virgin Lands}, so Kryukov could not possibly be the
author of {\it Quiet Flows the Don}.

\section{On the pointillistic processing of photographs}

Contemporary digital photography constructs images by means of a
large number of points, just as pointillist painters created
their canvasses. We have carried out a study based on digital
photographs.

We processed 24-bit files of images in the format $bmp$, which at
present is the optimal version from the standpoint of quality.
Each point in such a file is presented by 24 bits or 3 to 8
bytes. Thus each point is the mixture of three basic colors (red,
green, and blue), with one byte for each basic color, or 256
hues. The total number of colors equals
$256\times256\times256=16\,777\,216$ colors.

The processing was carried out as follows. From the image, we
obtained the matrix of colors: each point was represented in scale
16 (3 bytes). If this number is translated to the decimal scale,
we obtain what we consider to be the number of the color. In
simpler words, each color has its number (from 1 to
$16\,777\,216$) in this ordering, and our processing yields a
matrix of the numbers of the order for these colors. Further we
computed the occurrence frequency of each color and construct the
standard plots of the theoretical and experimental distributions,
and plotted the difference between these two magnitudes.

We chose several color pictures of very different types (see
figures 1--7). Five of them (``Computer graphics,'' ``A~flower,''
``New York by night,'' ``The tiger,'' and ``Mona Lisa'') were
rather well approximated, while two others (``Boyaryna Morozova''
and ``The island'') were poor approximations, the latter probably
because the little tree-covered island in the middle of the lake
shown on the picture does not give a uniform ``chaos.''

\begin{figure}
  \centering
  \includegraphics[viewport=0 0 1280 960,width=0.8\textwidth]{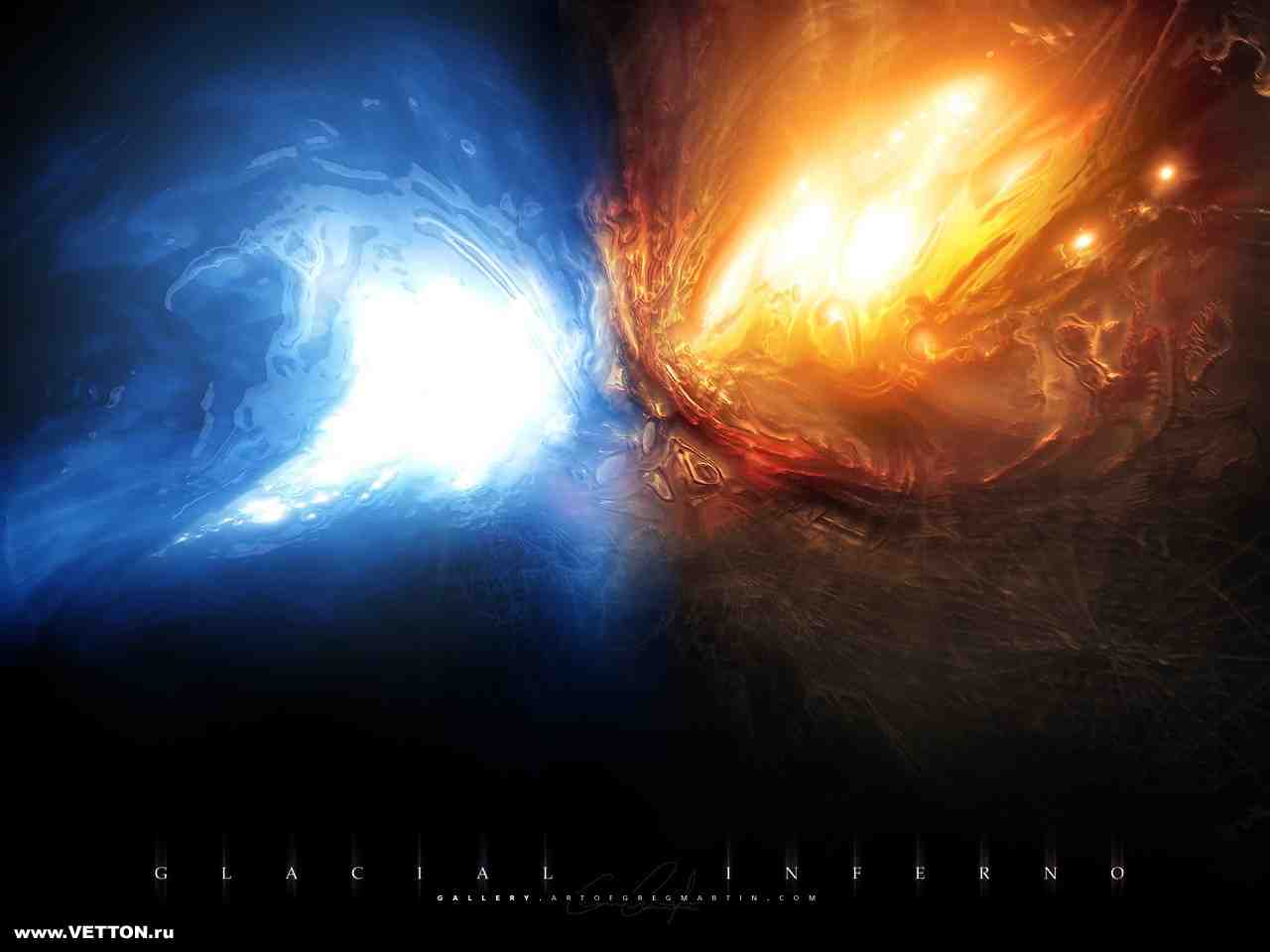}\\
  \includegraphics[width=0.5\textwidth]{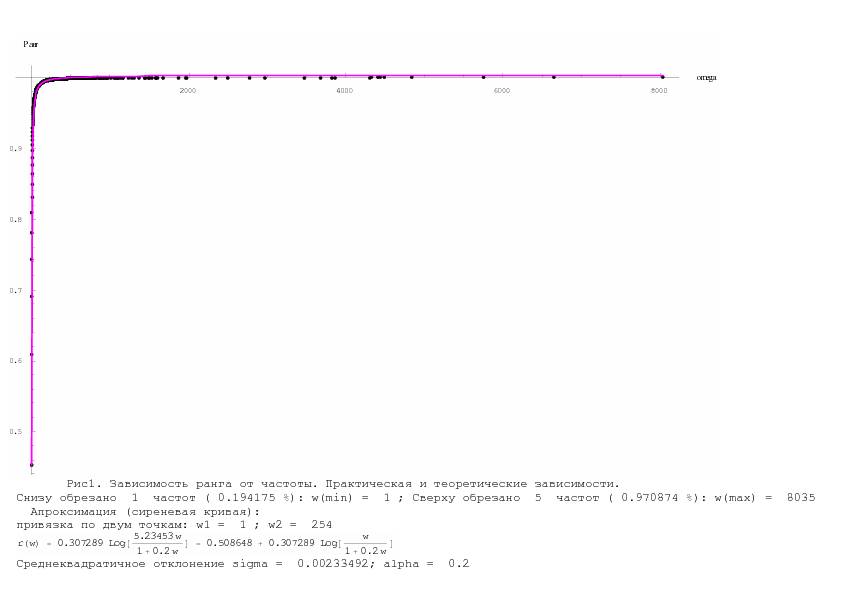}%
  \includegraphics[width=0.5\textwidth]{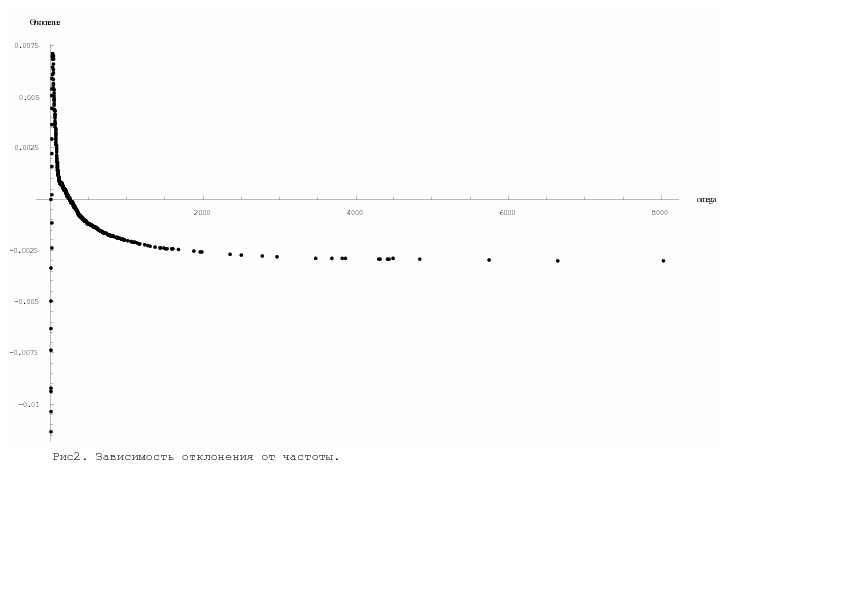}\\
  \caption{Top pane: image ``Computer graphics''
    (\protect\url{http://www.vetton.ru/walls/art/vetton_ru_705.jpg}).
    Bottom pane, left: the rank-frequency relation [circles: data;
    magenta line: least-squares fit $r(\omega) = (0.7576911 \pm
    0.012489081) + (0.1535386 \pm 0.007973972) \log(\omega/(1 +
    0.2\omega))$, where coefficients are given with 95\% confidence
    intervals], right: error of the rank-frequency
    relation.}
\end{figure}

\begin{figure}
  \centering
  \includegraphics[viewport=0 0 1152 864,width=0.8\textwidth]{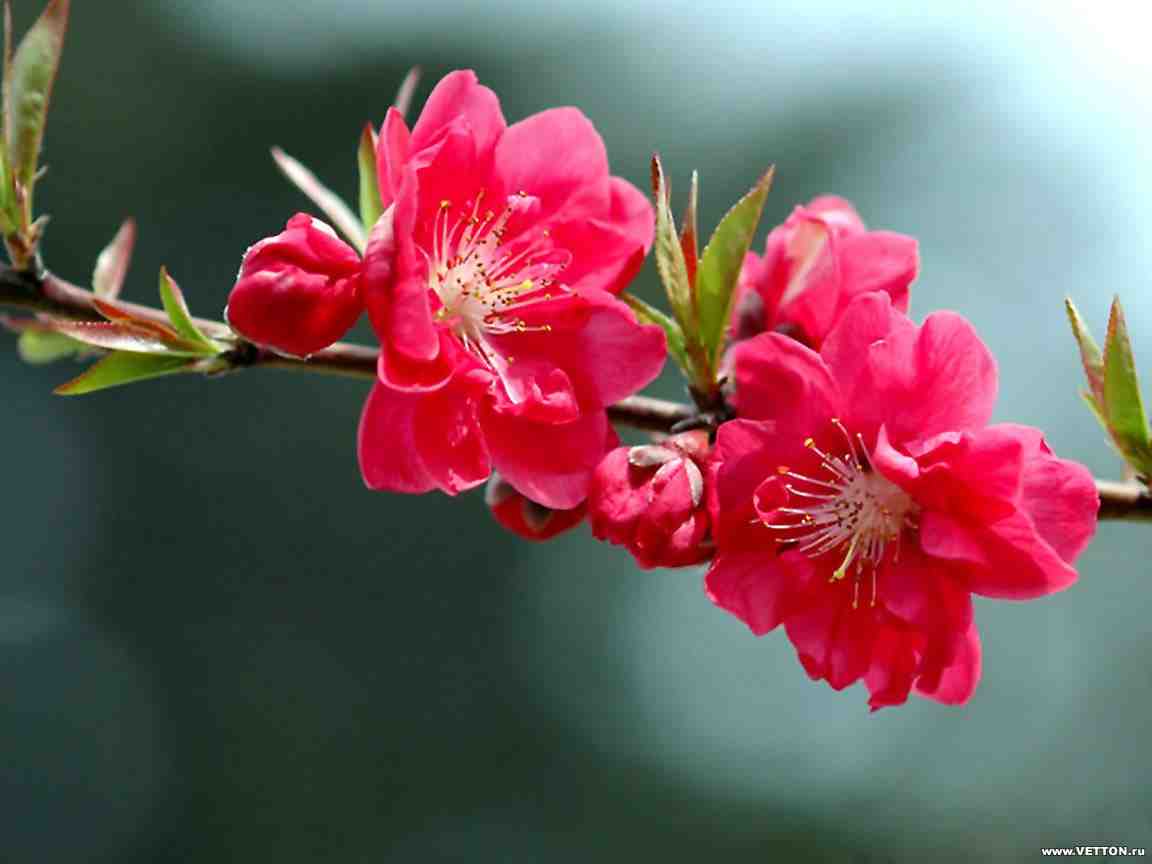}
  \includegraphics[width=0.5\textwidth]{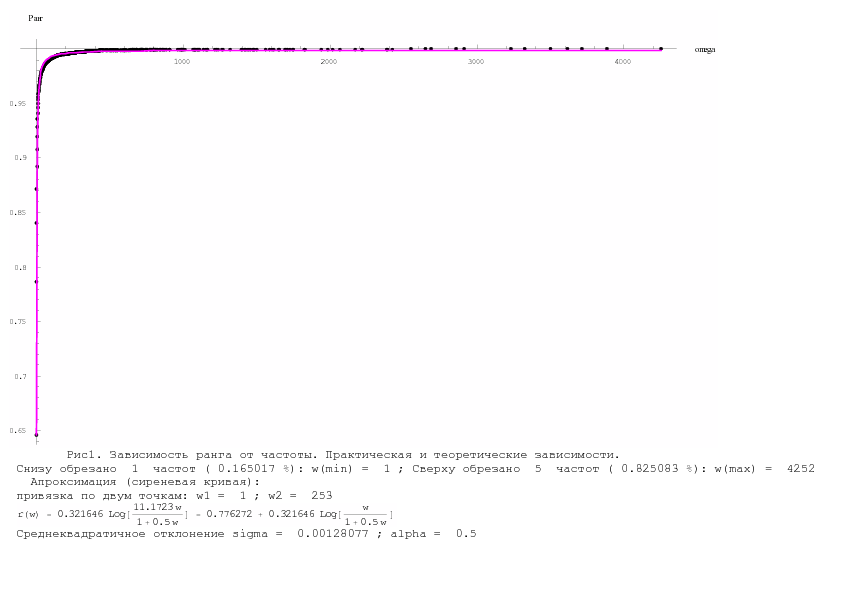}%
  \includegraphics[width=0.5\textwidth]{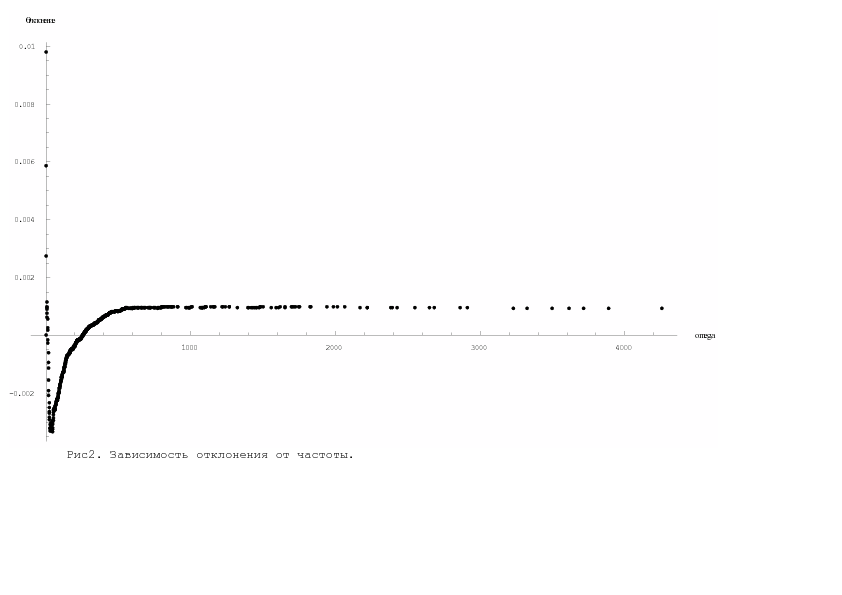}\\
  \caption{Top pane: image ``A~flower''
    (\protect\url{http://www.vetton.ru/walls/art/vetton_ru_175.jpg}).
    Bottom pane, left: the rank-frequency relation [circles: data;
    magenta line: least-squares fit $r(\omega) = (0.87733668 \pm
    0.008050773) + (0.07762449 \pm 0.005122996) \log(\omega/(1 +
    0.2\omega))$, where coefficients are given with 95\% confidence
    intervals], right: error of
    the rank-frequency relation.}
\end{figure}

\begin{figure}
  \centering
  \includegraphics[viewport=0 0 1600 1200,width=0.8\textwidth]{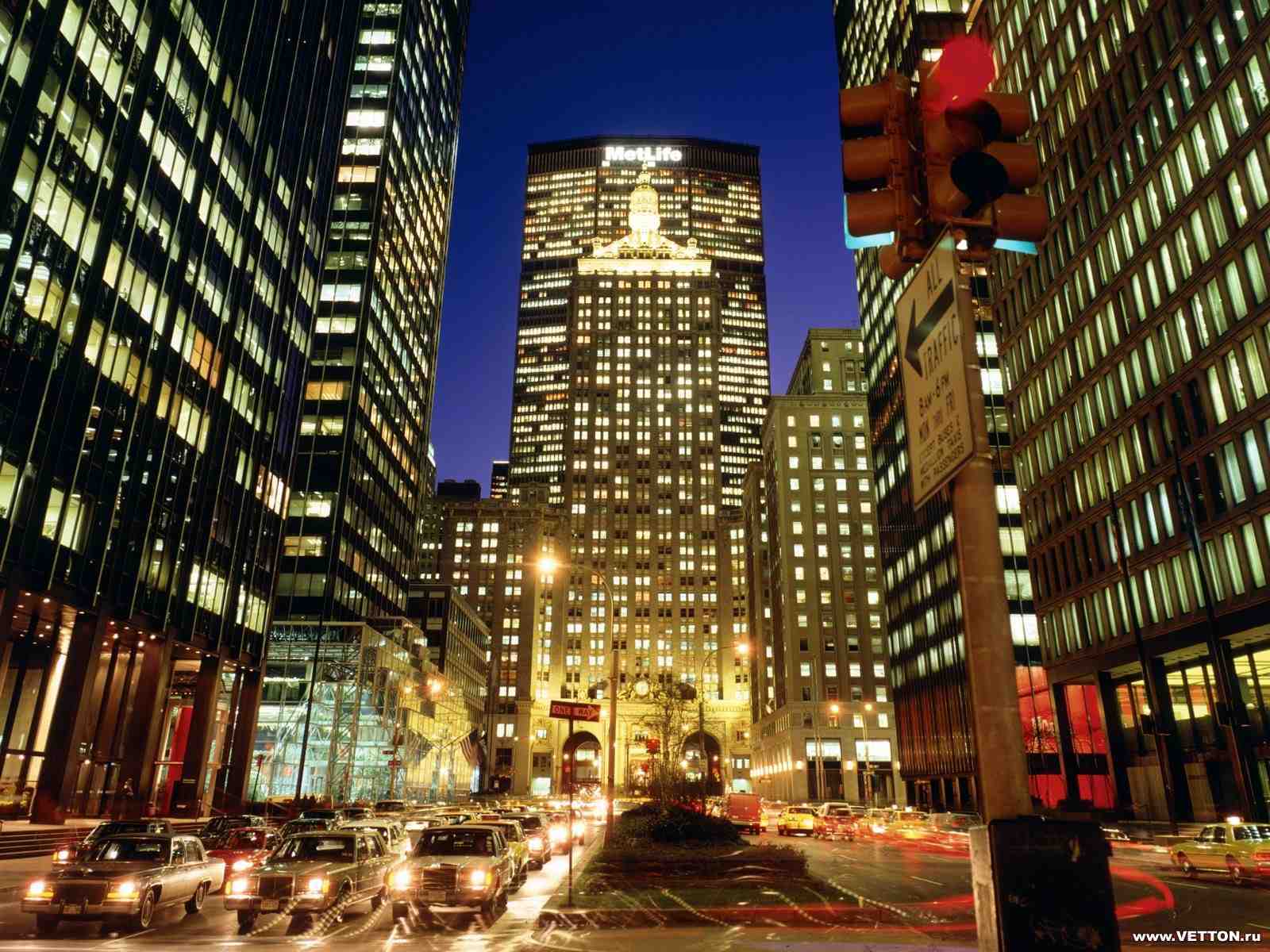}
  \includegraphics[width=0.5\textwidth]{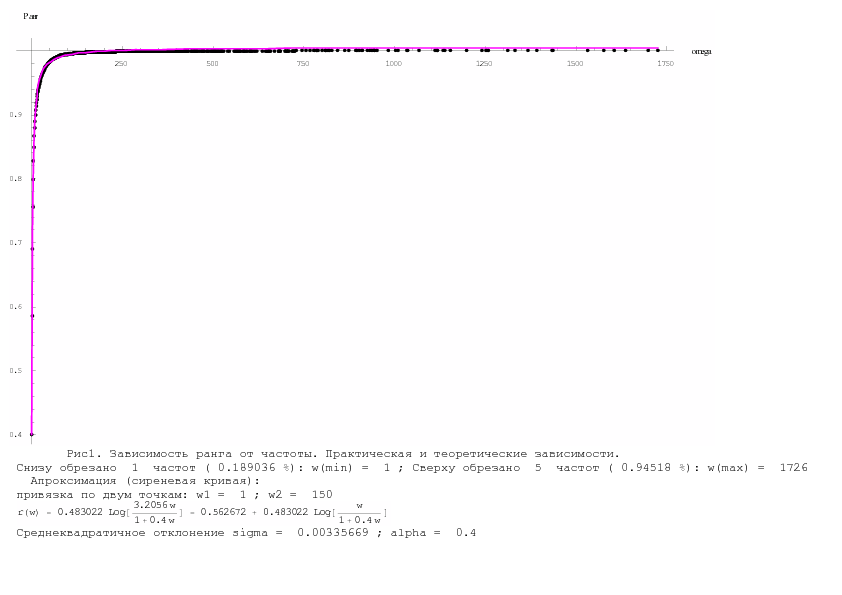}%
  \includegraphics[width=0.5\textwidth]{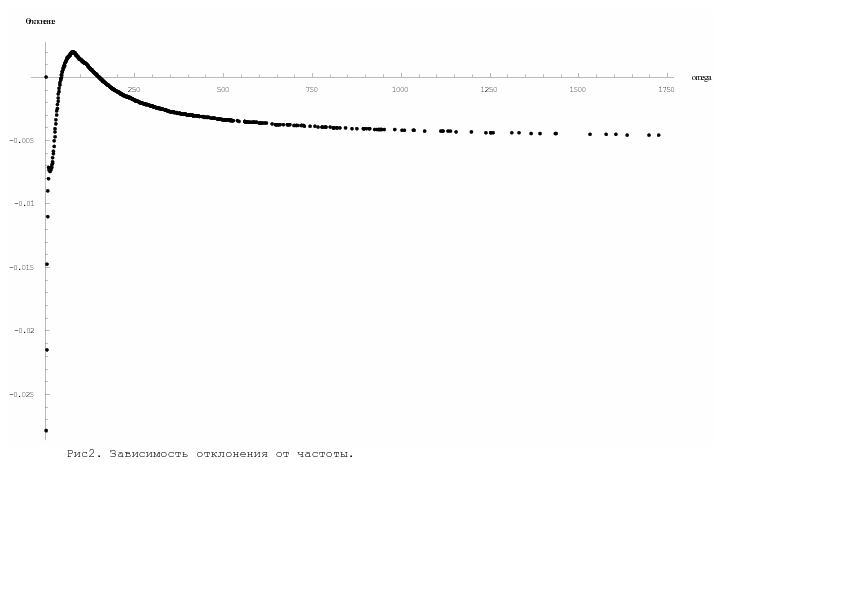}\\
  \caption{Top pane: image ``New York by night''
    (\protect\url{http://www.vetton.ru/walls/art/vetton_ru_231.jpg}).
    Bottom pane, left: the rank-frequency relation [circles: data;
    magenta line: least-squares fit $r(\omega) = (0.6807073 \pm
    0.01771724) + (0.2023832 \pm 0.01130678) \log(\omega/(1 +
    0.2\omega))$, where coefficients are given with 95\% confidence
    intervals], right: error of
    the rank-frequency relation.}
\end{figure}

\begin{figure}
  \centering
  \includegraphics[viewport=0 0 1114 834,width=0.8\textwidth]{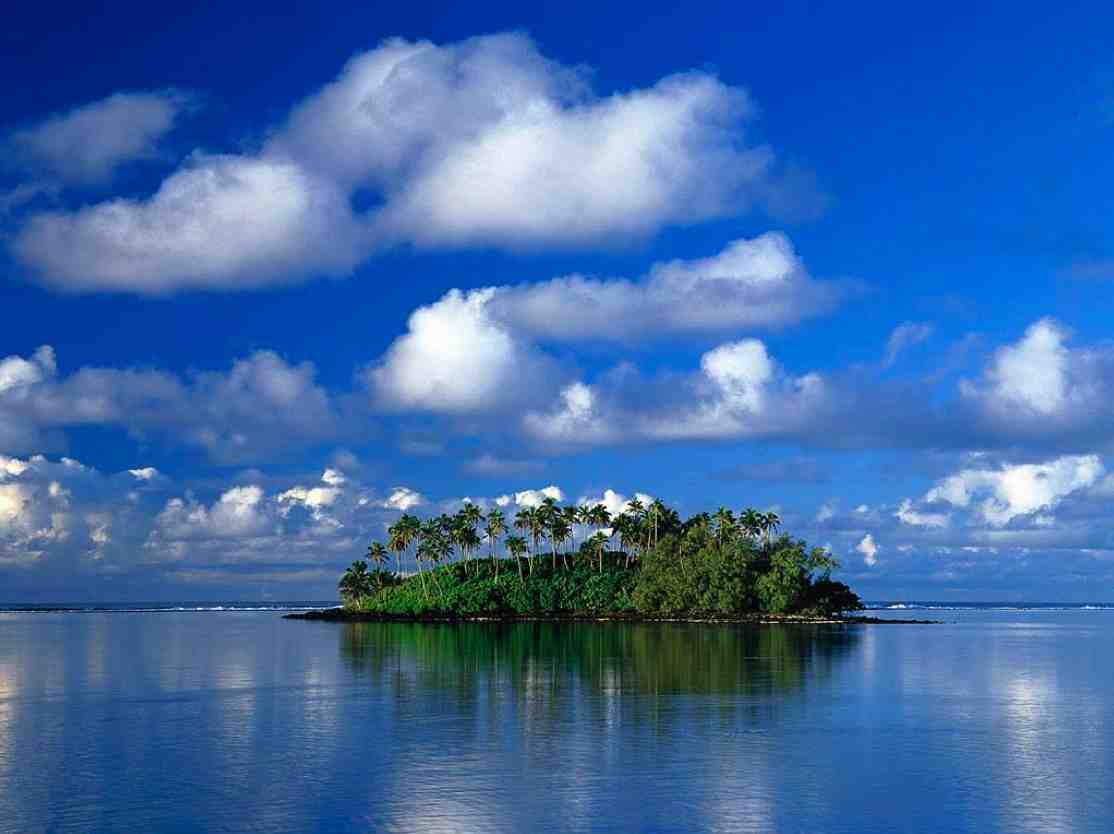}
  \includegraphics[width=0.5\textwidth]{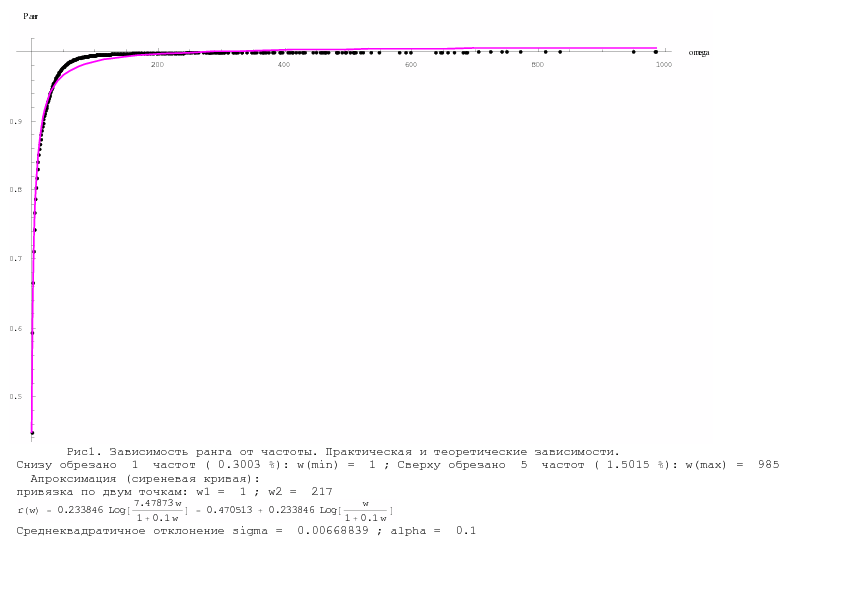}%
  \includegraphics[width=0.5\textwidth]{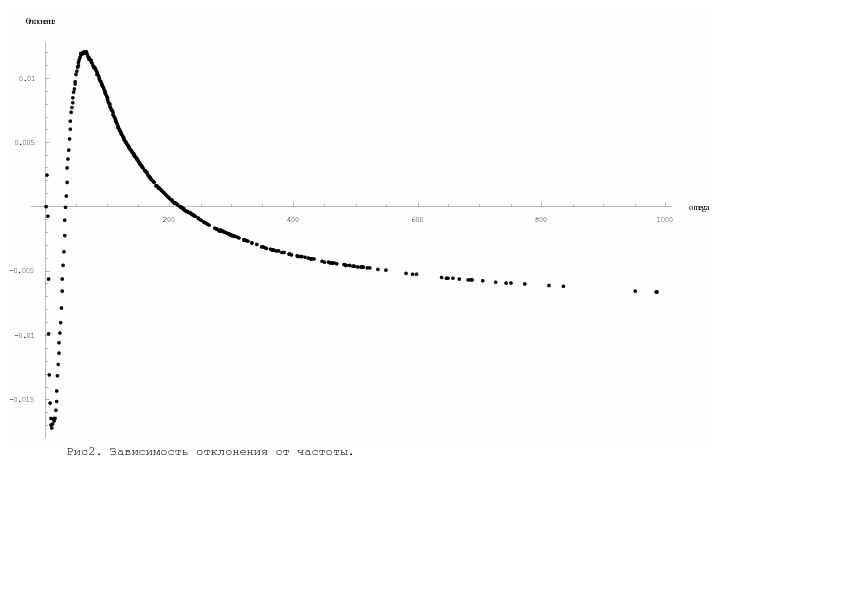}\\
  \caption{Top pane: image ``The island.''
    Bottom pane, left: the rank-frequency relation [circles: data;
    magenta line: least-squares fit $r(\omega) = (0.7673719 \pm
    0.013762098) + (0.1483364 \pm 0.008889888) \log(\omega/(1 +
    0.2\omega))$, where coefficients are given with 95\% confidence
    intervals], right: error of
    the rank-frequency relation.}
\end{figure}

\begin{figure}
  \centering
  \includegraphics[viewport=0 0 1600 1200,width=0.8\textwidth]{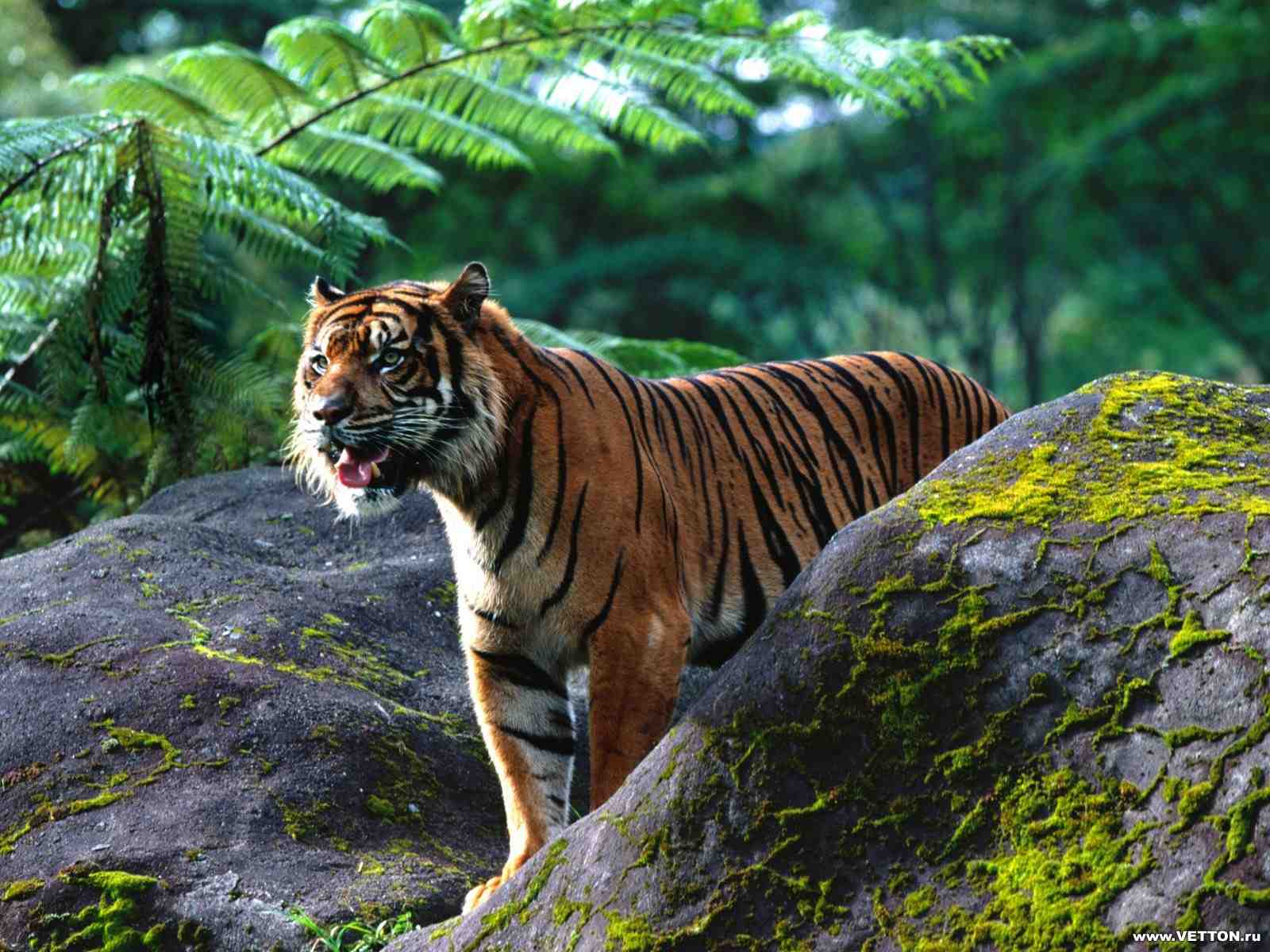}
  \includegraphics[width=0.5\textwidth]{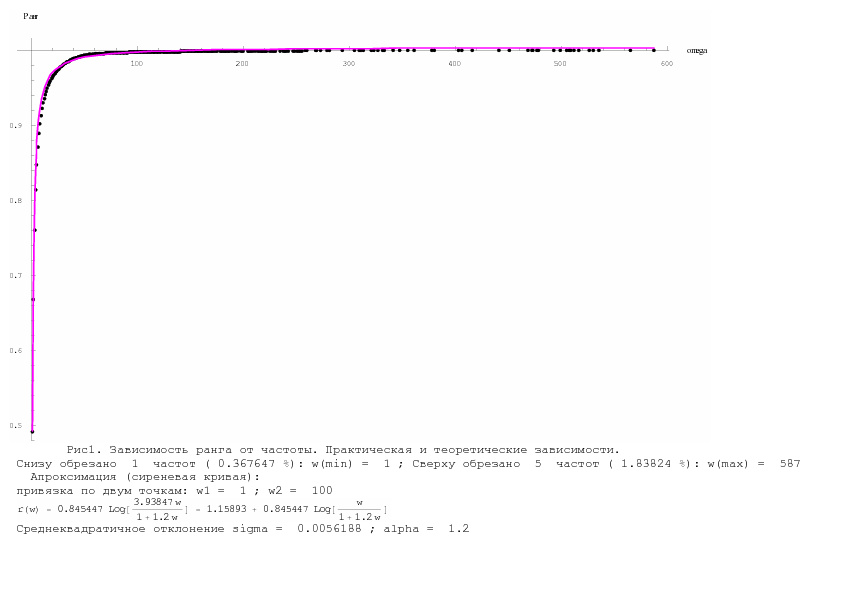}%
  \includegraphics[width=0.5\textwidth]{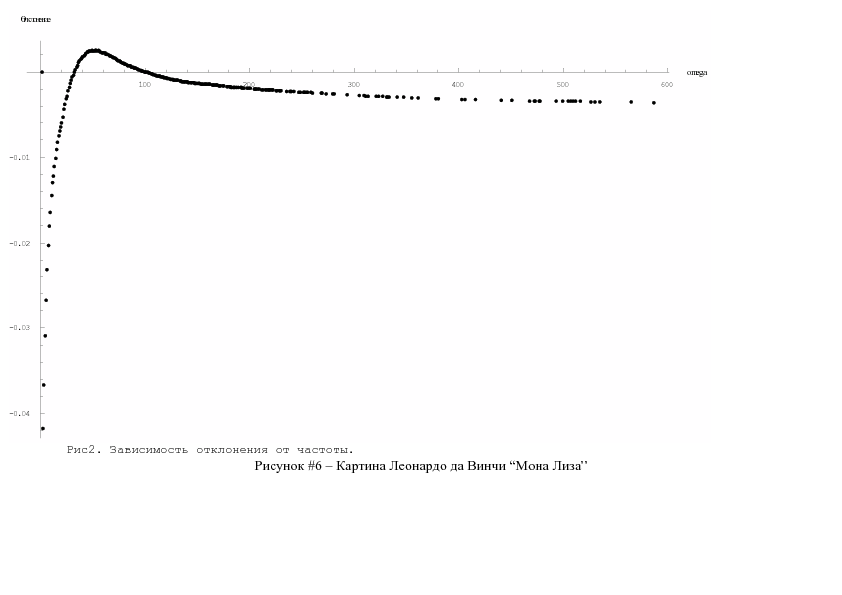}\\
  \caption{Top pane: image ``The tiger''
    (\protect\url{http://www.vetton.ru/walls/art/vetton_ru_331.jpg}).
    Bottom pane, left: the rank-frequency relation [circles: data;
    magenta line: least-squares fit $r(\omega) = (0.7555128 \pm
    0.01944519) + (0.1572145 \pm 0.01265441) \log(\omega/(1 +
    0.2\omega))$, where coefficients are given with 95\% confidence
    intervals], right: error of
    the rank-frequency relation.}
\end{figure}

\begin{figure}
  \centering
  \includegraphics[viewport=0 0 363 550,width=0.5\textwidth]{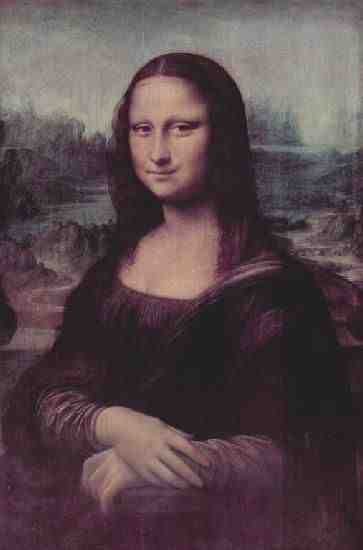}
  \includegraphics[width=0.5\textwidth]{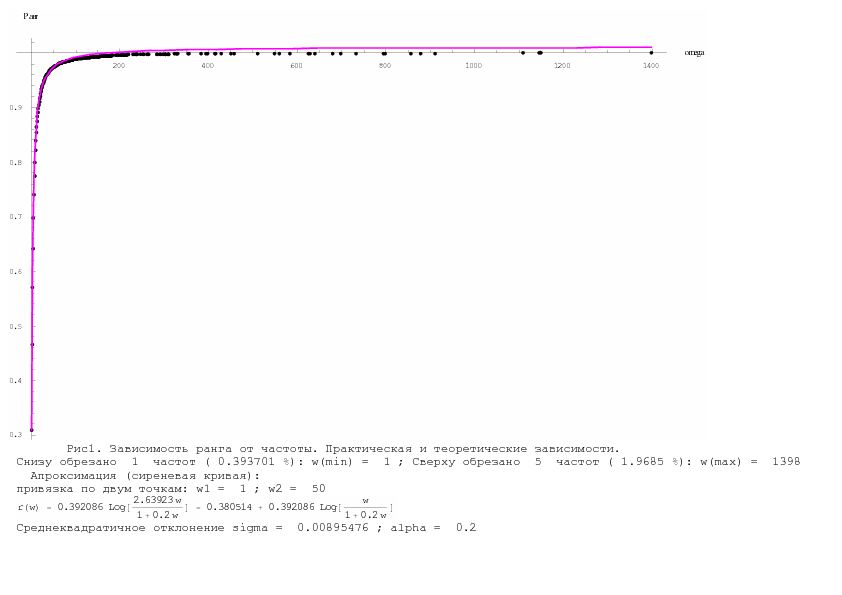}%
  \includegraphics[width=0.5\textwidth]{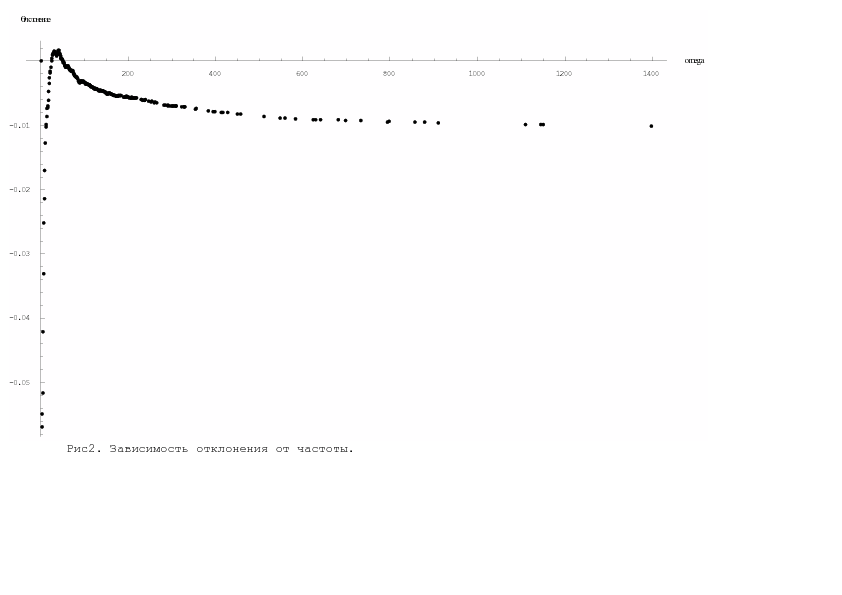}\\
  \caption{Top pane: image ``Mona Liza.''
    Bottom pane, left: the rank-frequency relation [circles: data;
    magenta line: least-squares fit $r(\omega) = (0.5541642 \pm
    0.02411542) + (0.2860553 \pm 0.01572469) \log(\omega/(1 +
    0.2\omega))$, where coefficients are given with 95\% confidence
    intervals], right: error of
    the rank-frequency relation.}
\end{figure}

\begin{figure}
  \centering
  \includegraphics[viewport=0 0 550 406,width=0.8\textwidth]{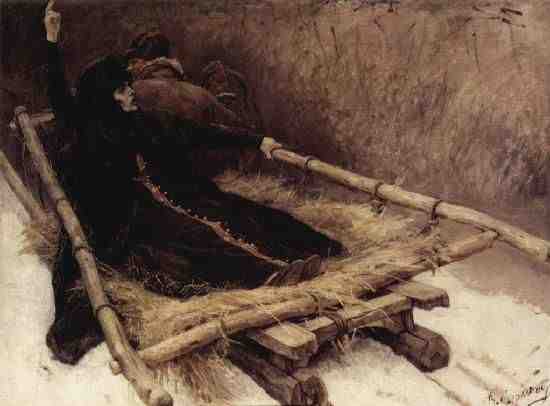}
  \includegraphics[width=0.5\textwidth]{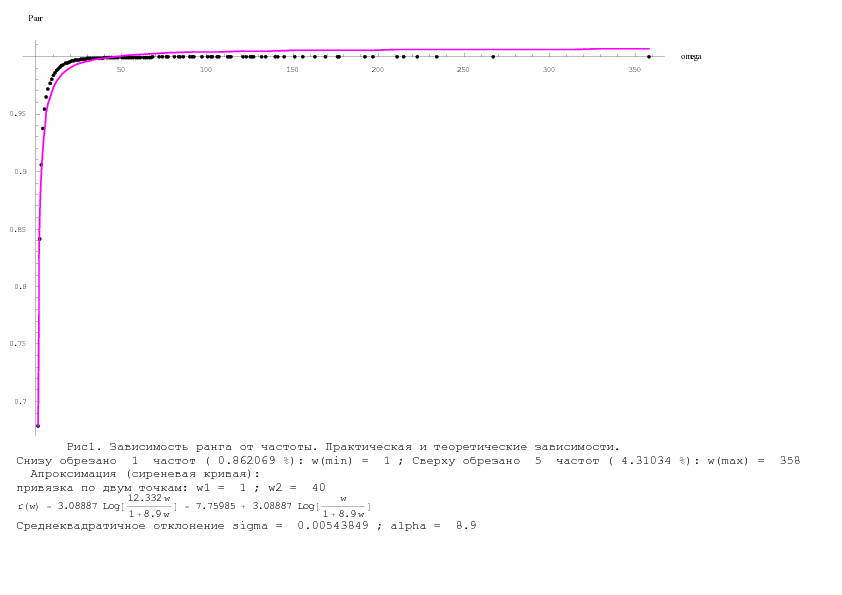}%
  \includegraphics[width=0.5\textwidth]{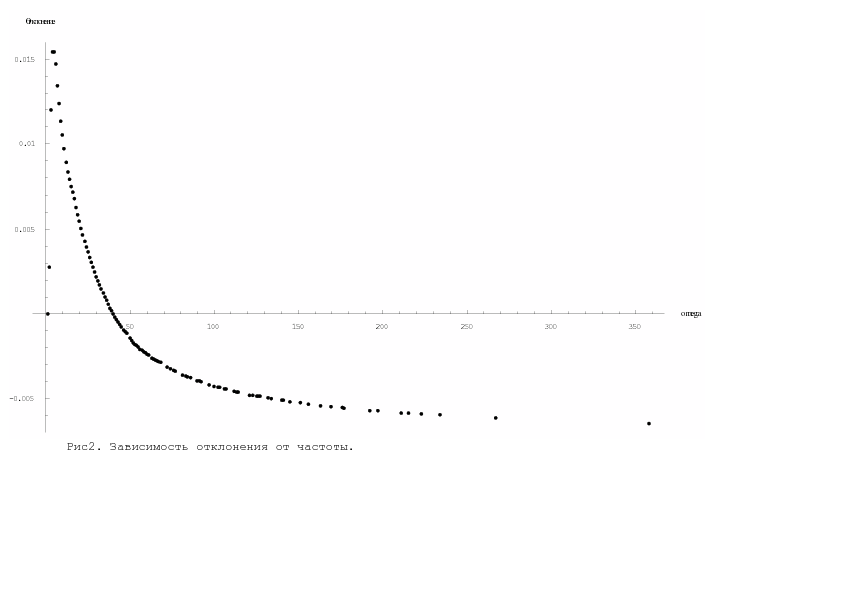}\\
  \caption{Top pane: image ``Boyarynya Morozova.''
    Bottom pane, left: the rank-frequency relation [circles: data;
    magenta line: least-squares fit $r(\omega) = (0.85155937 \pm
    0.02285417) + (0.09873425 \pm 0.01552277) \log(\omega/(1 +
    0.2\omega))$, where coefficients are given with 95\% confidence
    intervals], right: error of
    the rank-frequency relation.}
\end{figure}

The pictures were processed by N.~Marchenko, who also learned how
to ``sew in'' texts into pictures by steganographic methods. He
took the color photograph of the tiger and ``sewed into it'' the
text of Shakespeare's {\it King Richard the Third} containing
$190\,000$ symbols. The size of the file was not changed and the
pictures of the tiger with or without King Richard were visually
indistinguishable.

Further, one may find the dependence of the rank on the frequency
for the modified picture with respect to the central line. To do
this, one should take all pairs of points symmetric with respect
to the central line, and then take the mean value of the colors
$R=(R_1+R_2)/2$, $G=(G_1+G_2)/2$, $B=(B_1+B_2)/2$ (thus in fact
the upper half of the picture is rotated by 180 degrees in space
and superimposed on the lower one). Then one may plot the
dependence of rank on frequency as before.

After such a modification, the dictionary becomes $1.5$ times
larger, and the maximal frequency decreases, and decreases
significantly. This means that the picture has become much too
gaudy (the distribution of colors has become more chaotic). The
plot shows the degree of asymmetry. The inclusion of new objets
must not breach this asymmetry.

\section{Architecture. How to built within an existing ensemble}

Synergetic studies show that human beings tend to acquire the
opinions of the majority. It is known that the soldiers of the
Russian army sent by the Commander-in-Chief to put an end to the
February revolution in Saint Petersburg in 1917 were immediately
imbued with the revolutionary spirit and took the side of the
rebels.

G.~Haken stressed the importance of a phenomenon which in physics
led to the creation of laser emission. One of the authors of the
present paper discovered, already in 1958, a remarkable
phenomenon. If one slightly bends the parallel infinite mirrors
in a laser, then a single mode laser emission occurs: the rays
will not run away to infinity by diffraction, but will remain
inside the infinite plate. Thus, from a chaotic bundle of rays in
the open laser, a monochromatic wave is created, it remains
within the open resonator, increases and, instead of diffusing
according to the properties of waves all along the resonator, it
remains in its ``finite'' part.

At the time it was impossible to construct such a monochromatic
wave in narrow reflecting surfaces. At present this can probably
be achieved by means of nanotechnologies.

Such a resonance phenomenon in human society and nature is still
unexplained as a phase transition of the zeroth kind and requires
additional study. It may be related to the sharpening regimes
which S.\,P.~Kurdyumov correlates with synergy problems.

However, it appears that the instrument best suited for falling
into resonance is actually the human being. The spectator laughs
or cries watching a film. Hostages are contaminated by the
ideology of the terrorists that keep them in captivity (the
Brussels syndrome). How people progressively become rhinoceroses
(meaning fascists), even when they are least likely to be
influenced by nazi ideology, is brilliantly shown by Ionesco in
his play {\it The Rhinoceroses}. But if arians were easily
converted to this ideology, nonarians were unable to become
arians and therefore fascism was doomed from the outset.

The situation is quite different when invaders impose their
religion and converting to it suffices not only to survive but to
acquire equal rights. In that case, the newly converted sometimes
become more fanatically zealous in their religion than the
invaders.

B.\,F.~Porshnev put forward the following paradoxal point of view
on the appearance and development of mankind. Men at first were
parasites attracted to the carnivores, like Maugli, and easily got
used to their food and their voices. And while the wolf cub's
voice would change as it grew up, and it was rejected by its
mother, a human could continue to produce the same sounds, and
the she-wolf (or she-lion or other carnivore) would continue
feeding the human. This easy adaptation to different kinds of
foods, to different languages is characteristic of humans.

This point of view contradicts the usual one, according to which
mankind created domestic animals. But from Porshnev's point of
view, humans first were the parasites of animals which they
subsequently enslaved, as sometimes happens with people in
everyday life.

In nature, the self-organization of animals takes place slowly,
except perhaps during catastrophes. For people, we have the
expression ``become part of a collective,'' ``part of a team.''
Humans become part of nature, which organizes itself. This
determines the ultimate harmony and beauty of nature. Humans relax
simply by observing it.

In the same way, ancient cities with their specific architecture
fit in, blend into the surrounding landscape. This feeling
increases if the observer is familiar with the history of the
city. For example, after visiting York in England, a person
becomes ready to feel the Shakespearian plays related to the
period of the War of the Roses. The observer enters the
atmosphere of the Middle Ages and is involuntarily penetrated by
its spirit. Here everything is clear.

But if, inside an already existing ensemble of a city, a new
architectural construction is added, how can one make it blend
into the environment, self-organize itself into it, resonate with
it? Clearly the surrounding nature and the already existing
architectural ensemble will not adapt to the new construction
(will not self-organize itself together with the new
construction).

We know that Le Corbusier proposed to tear down Paris completely
and replace it by new buildings from his projects.

Of course, we have the option of changing the ensemble or the
surrounding landscape, but nature by itself will not resonate.

How could one verify if the Eiffel tower would blend into the old
Paris? Did the glass pyramids blend with the Louvre? The answer is
 --- yes, they did. Did the Kremlin Palace blend into the Kremlin
ensemble? The Rossiya Hotel? One can photograph the ensemble and
study the plots corresponding to the photographs with or without
the palace; how will the ``self-organization'' plots change?

In our opinion, it is precisely this objective criterion which
must determine whether or not a new building spoils the ensemble
into which it is enclosed.

Whether or not such an addition enhances the effect of the
ensemble is another question, it depends on the talent of the
architect. But the objective verification of how it blends into
the self-organized harmony is, of course, useful.

We used this approach when we carried out a practical experiment in
the construction of out-of-town cottages. The houses were constructed
in a forest, and our aim was to blend the structures into the
``ensemble'' of the forest, without touching it (Fig.~\ref{fig:forest}
on p.~\pageref{fig:forest}). This involved, first, making apertures
for the roots of the surrounding trees (the construction machines did
not touch the trees).  Secondly, we~photographed the structures only
from the sides that were visible through the forest. Thirdly, we took
the photographs at different hours of the day (like Claude Monet, who
painted the same bales of hay at different hours) and at different
times of the year. Thus we obtained, as in the analysis of {\it Quiet
  Flows the Don}, a sufficiently wide spectrum of values for the
parameters $\beta$ and $\sigma$.

\begin{figure}[t]
  \centering
  \includegraphics[width=0.8\textwidth]{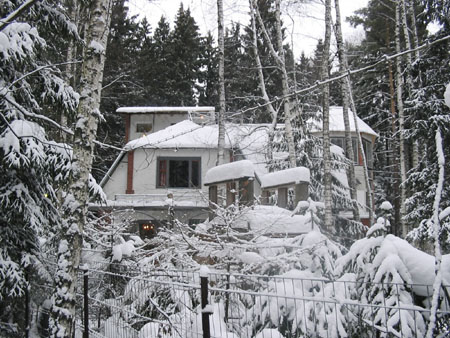}\\
  \includegraphics[width=0.4\textwidth]{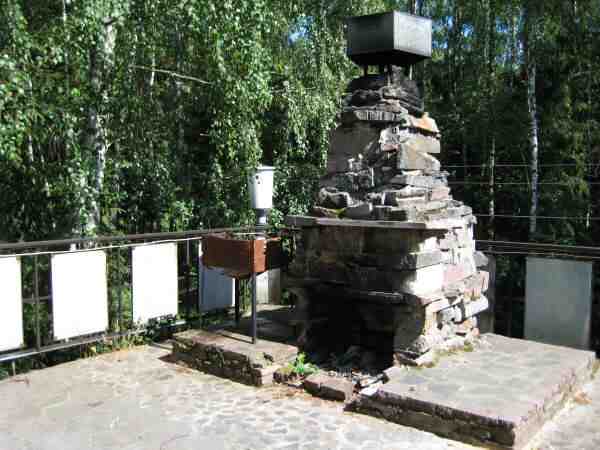}
  \includegraphics[width=0.4\textwidth]{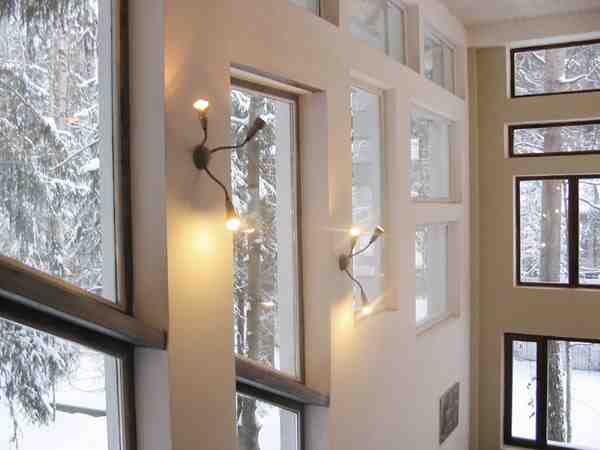}
  \caption{Images of a forest cottage.}
  \label{fig:forest}
\end{figure}

Therefore, from the point of view of steganography, this need not
be much of a problem. However, from the viewpoint of architecture,
since one can never look at a building as a whole, since it is
visible only from the open sides, this is a complicated approach.
It is not surprising that in an article in the newspaper
``Culture'' (formerly ``Soviet Culture'') describing one of the
cottages, entitled ``Windows of freedom and growth,'' the title
was followed by the subtitle ``Academician Maslov broke all the
rules of architecture, and not in vain.''

The computer allows to construct and study the picture before the
object is constructed and photographed. Experimenting with a
computer, we often remembered the words of Picasso: ``I~search ---
and I find.'' But this is not enough. In our practice, it turned
out that it was sometimes necessary to break down parts of the
constructions already made.

At the present time, numerous cottages have been constructed near
Moscow --- this is a trait of our times. Many years ago, when the
scientific town of Akademgorodok was being constructed near
Novosibirsk, academicians Lavrentiev, Sobolev, and Khristianovich
personally headed the construction and succeeded in saving a part
of the forest in the middle of the new town. For the creative
processes of research scientists, this was very important. It was
also important for the teaching staff: after a lecture, a
professor would walk through the forest and get rid of the
``inspiration'' needed to lecture, would relax. When one of the
construction workers chopped down a small fir tree for Christmas,
Lavretiev immediately threw him out of the city.

The forest has the same beneficial effect on composers, writers,
and stage directors. A businessman or a political figure,
returning to his out of town house after a hard day, can get rid
of the stress, change air and atmosphere, modify the style of his
office.

At the present, special computer software is used in the design
and construction of new buildings. But this is not enough. The
construction must fit into the ``collective.'' At the same time,
the construction should be sufficiently modern, may even be close
to the high tech style.

In our own practical construction, we were forced to modify the
already constructed object, because we saw the drawbacks of the
software that we used. It is necessary to correct
``architectural'' software so as to avoid such expensive mistakes.

The well-known architect F.Hundertwasser wrote: ``People think
that a building is its walls, but actually --- it's the windows.

Hundertwasser's remark about windows is very relevant to our
times. A person inside a house must feel like someone in a
bathiscape in the underwater world. He must see and feel the
exterior ensemble even when he is taking a shower.

But it is hard to agree with Hundertwasser when he claims that
there are no straight lines in nature, and hence they should not
appear in architecture. In nature, even in a mixed forest,
straight lines do appear: light rays, falling through the leaves
and producing shadows (compare, in painting, with the {\it
luchism} theory of M.~Larionov). Symmetry also appears, albeit
not too strict: reflections in water, the inner symmetry of
flowers, leaves, living organisms. There is also the fractal
structure of snowflakes. All this must be reflected in the
architecture of buildings that are to fit in the ensembles of
nature.

To such a complex approach, one can apply the mathematical notion
of ``complexity'': it is~inherent to nature, and it must be
applied in architecture.Thus, the architecture of buildings in
a~cone forest, in which fir trees stand like ``a shot from a
rifle,'' must differ from that in a mixed forest.

The relationship between mathematics and architecture has been
noted long ago, especially in the architectural masterpieces of
antiquity. Besides obeying the golden section rule, Greek temples
blend harmoniously with the surrounding nature: the rocky slopes
of the Acropolis in Athens, the Poseidon temple in Peloponnesus.
Especially when they are open to the sea, when the mathematical
harmony and rigor, the geometry of the doric columns,
involuntarily bring to mind the name of Archimedes.

Architectural geometry is especially expressive with the
background of a desert, as in the Sahara, where the Roman city of
Sbaitl is situated.

In Istanbul, in Kairuan, in muslim cities with their tradition of
covering the body and faces of wo\-men by clothing, this geometry
has no limits. It is not surprising that the mosques of Istanbul,
built on the basis of ancient temples, columns were turned upside
down, with the capitols at ground~level.

In Samarkand, we see a different kind of beauty, another geometry
--- the beauty of fractals. In Mexico architecture is related to
rocks.

The relationship to rocks is also visible in gothic architecture,
especially in certain castles. Here the geometry is also clearly
expressed.

Sayings about semiotics in art (see \cite{29}), relate especially
well to architecture, where it is possible to identify and
distinguish a set of signs.

In our approach, we look at the problem from a different angle,
when the architectural signs in a complex structure are numerous
and the structure is placed inside a very complex ensemble, and
therefore the risk of destroying the harmony given to us by
nature is high. It suffices to drive through some of the ugly
settlements near Moscow to appreciate how serious the problem is.
Or to visit the unique vacation center ``Krainka'' in the Tula
region, where a magnificent old park, with trees situated so that
their fallen leaves form a beautiful rug in autumn, is
``embellished'' by sculptures of half-naked javelin throwers. It
suffices to photograph these ``sculptures,'' then delete them
from the photograph, and compare the corresponding plots to see
that these sculptures are incompatible with the harmony of the
park. If we take a hammer and, following Rodin, remove all which
is not needed, we may be left with little pieces of stone that
will fit into the park. M.~Voloshin used to repeat that a
sculpture only acquires its final form after it is thrown down
from a mountain.

Finally, a last general remark about architecture. No one today
seriously considers adding arms to the Venus de Milo or restoring
the missing pieces of the Perham altar. But in Russia many think
that one can take an architectural structure and move it away from
the place where it was constructed (as this was done in Suzdal),
paint something over, renovate something, give the cupolas a
shine. We propose, if such actions be undertaken, to first check
the result by means of computations similar to those described in
the present paper.

Returning to architecture in the forest, note that the structure
of a house in the forest must be such that one will never tire of
it, never be satiated with it. Your house has to be like a drug.
The~desire to return there must be like the nostalgia for one's
birthplace. You must feel attracted to your house, so that when
you come in and light the chimney, you can sigh and say: finally,
I am at home.

A house in the forest must not have any distances. You are in the
forest. You have forgotten about the tedium of everyday life.
Having plunged into this communion with nature for only a half
hour, you should be able to get rid of the weight of your
problems.

Different illuminations of the house, like different stage sets,
must correspond to the mood that you want your close ones and
yourself to feel. Thus, in Pitoev's theater in Paris, at a certain
time, the change of stage settings was replaced by changes in
illumination. The illumination of the house, when you stand in the
backyard, must be transparent, enhancing the beauty of the forest,
so that you will want to cry out --- ``how beautiful!'' And that
is what architecture in a forest is about.

One can readily see that an asphalt road passing through a forest
is a discord. Even in parks (e.g., in Versailles) the paths
should not be covered by concrete. And all the more, if we want
to preserve the heavenly chaos of a wild mixed forest.
Cobblestones and gravel cost a lot more than asphalt, but the
health of people related with what they see and smell is worth a
lot more.

The beauty of ``the fractal'' is very well described by the
outstanding Russian painter and graphic artist
M.\,V.~Dobuzhinsky, who really knew how to find sharp vantage
points for his drawings of architectural fractals, viewpoints
from which he obtained remarkable combinations of roofs, walls,
columns, and other structures, briefly, of parts of an ensemble.
In his ``Recollections'' he wrote: ``All~this I had already seen
as a child on maps \dots and I started looking even more
attentively at various geographic lines and silhouettes, noticing
their remarkable elegance and, so to speak, of their organic
structure. I was also very fond of finding unexpected repetitions
of certain shore lines and contours of land, seas, lakes, and
bays \dots I was also surprised by the repetition of some
geographic outlines (e.g., the Peloponnes and the Galipoli
trident, the Dutch Zeedersee and the Caspian bay of Karabogas,
similar sand spits of Memel in the Baltic and Kinburn in
Crimea)'' \cite{30}, p.~99. It is well known that shorelines are
fractals. The relationship between fractals and the formulas
obtained above is characterized in \cite{6}. The beauty of
asymmetry, of divine chaos, of self-organizing chaos, obeying
explicit laws, must be underlined, additionally stressed by the
geometry of the added architectural structure, by the
illumination. At least this should be attempted. As to the
mathematical formulas, they can only show that we have not
destroyed the already existing~harmony.
\smallskip

The authors would like to express their gratitude to A.~Churkin
and N.~Marchenko for help in computer calculations.


\begin{thebibliography}{99}
\bibitem{00}
V.\,P.~Maslov and T.\,V.~Maslova, ``Synergetics and
Architecture,'' Russ. J. Math. Phys. {\bf 15} (1), 102--121 (2008)
\bibitem{3}
A.\,N.~Kolmogorov, {\it Collected Works in Mathematics and
Mechanics} (1985), pp.~136--137
\bibitem{9}
V.\,P.~Maslov, ``Nonlinear Mean
in Economics,'' Mat. Zametki {\bf 78} (3), 377--395  (2005)
[Math. Notes {\bf 78}, 807--813 (2005)]
\bibitem{10}
O.~Viro, {\it Dequantization of Real Algebraic
Geometry on Logarithmic. Texts for Students} (2001);\newline {\tt
http://www.pdmi.ras.ru/~olegviro/dequant}
\bibitem{11}
M.~van Manen and D.~Siersma, ``Power Diagrams and
Their Applications,'' {\tt arXiv:math/05058037,} Aug~2, 2005
\bibitem{12}
D.~Alessandrini, ``Logarithmic Limit Sets of Real Semi-Algebraic
Sets,'' {\tt arXiv:0707.0845 [math.AG] v1} 5 Jul 2007
\bibitem{13}
P.~Maslov, ``Phase Transitions of the Zeroth
Kind,'' Mat. Zametki {\bf 76} (5), 749--761  (2004)
\bibitem{14} V.\,P.~Maslov, ``Quasistable Economics and Its
Relationship with the Thermodynamics of Superfluid Liquids.
Default as a Phase Transition of Zeroth Kind. I'' Review of
Applied and Industrial Mathematics, Section ``Mathematical
Methods in Economics'' {\bf 11} (11), 630--732 (2004); II {\bf
12} (1), 3--40 (2005)
\bibitem{1}
G.~Haken, ``Self-Organizing Society,'' in
{\it The Future of Russia in the Mirror of Synergetics}
(KomKniga, Moscow, 2006), pp.~194--208
\bibitem{8}
C.~Maincer, ``Complexity Challenges Us in the 21st Century:
Dynamics and Self-Organization in the Age of Globalization'' in
{\it The Future of Russia in the Mirror of Synergetics}
(KomKniga, Moscow, 2006), pp.~209--226
\bibitem{4}
V.\,P.~Maslov, ``Expertise and Experiment,'' Novy Mir {\bf 1},
243--252 (1991)
\bibitem{5}
V.\,P.~Maslov, {\it Quantum Economics} (Nauka, Moscow, 2006) [in
Russian]
\bibitem{6}
V.\,P.~Maslov, ``Revision of Probability Theory from the
Point of View of Quantum Statistics,'' Russ. J. Math. Phys. {\bf
14} (1), 66--95 (2007)
\bibitem{7}
V.\,P.~Maslov, ``The Lack-of-Preference Law and the Corresponding
Distributions in Frequency Probability Theory,'' Mat. Zametki
{\bf 80} (2), 220--230  (2006) [Math. Notes {\bf 80}, 214--223
(2006)]
\bibitem{16}
V.\,P.~Maslov, ``On the Principle of Increasing Complexity in
Portfolio Formation on the Stock Exchange,'' Dokl. Acad. Nauk
{\bf 404} (4), (2005) [Doklady Mathematics {\bf 72} (2), 718--722
(2005)]
\bibitem{15}
L.\,D.~Landau and E.\,M.~Lifshits, {\it Statistical
Physics} (Nauka, Moscow, 1976)

\bibitem{17}
V.\,V.~Vyugin and V.\,P.~Maslov, ``On Extremal Relations between
Additive Loss Functions and Kolmogorov Complexity,'' Probl. Inf.
Transm. {\bf 39} (4), 71--87 (2003)
\bibitem{18}
A.\,V.~Podlasov, ``The Theory of Self-Organizing Criticality as
the Science of Complexity,'' in {\it The Future of Applied
Mathematics. Lectures for Young Researchers,} Ed. by
G.\,G.~Malinetskii (Editorial URSS, Moscow, 2005), pp.~404--426
\bibitem{19}
A.\,T.~Fomenko and T.\,G.~Fomenko, ``An Invariant of Authors of
Russian Literary Texts,'' in {\it Methods of Quantitative
Analysis of Texts from Narrative Sources} (USSR Academy of
Sciences, Institute of the History of the USSR, Moscow, 1983),
pp.~86--109
\bibitem{26}
V.\,P.~Maslov and T.\,V.~Maslova, ``On the Zipf Law
and Rank Distributions in Linguistics and Semiotics,'' Mat.
Zametki {\bf 80} (5), 718--732 (2006) [Math. Notes {\bf 80},
806--813 (2006)]
\bibitem{28}
T.\,V.~Maslova, ``On the Specification of Zipf's Law for Frequency
Dictionaries,'' NTI, Series 2, no.~11, 37--44 (2006)
\bibitem{21}
B.~Mandelbrot, {\it The Fractal Geometry of Nature}   
(Istitute of Computer Studies, Moscow, 2002) [in Russian]
\bibitem{22}
 M.\,V.~Arapov, E.\,N.~Efimova, and Yu.\,A.~Shreider,
``On the Meaning of Tank Distributions,'' NTI, Series 2, no.~2,
9--20 (1975)
\bibitem{20}
{\it British National Corpus}, {\tt
ftp.itri.bton.ac.uk/pub/bnc}
\bibitem{23}
{\it Mathematical Encyclopedic Dictionary} (Moscow, 1988)
\bibitem{25}
B.\,M.~Malyutov, ``Survey of Methods and Examples of the
Attribution of Texts,'' Review of Applied and Industrial
Mathematics {\bf 12} (1), 41--77 (2005)
\bibitem{24}
V.\,P.~Maslov, ``Quantum Linguistic Statistics,'' Russ. J. Math.
Phys. {\bf 13} (3), 315--325 (2006)
\bibitem{arX_Neg_Dim}
V.\,P.~Maslov, ``Negative Dimension in General and Asymptotic
Topology,'' {\tt arXiv:math.GM/0612543,} Dec~19, 2006
\bibitem{arX_Hole_Dim}
V.\,P.~Maslov, ``Dimension of Holes and High-Temperature
Condensate in Bose-Einstein Statistics,'' {\tt
arXiv:physics/0612182,} Dec~22, 2006
\bibitem{SecondDequant}
V.\,P.~Maslov, ``Secondary Dequantization in Algebraic and
Tpopical Qeometry,'' Mat. Zametki {\bf 82} (6), 953--954 (2007)
[Math. Notes {\bf 82}, 860--862 (2007)]
\bibitem{29}
V.\,V.~Ivanov, {\it
Essays on the History of Semiotics in the USSR} (Nauka, Moscow,
1976) [in Russian]
\bibitem{30}
M.\,V.~Dobuzhinsky, {\it Recollections} (Nauka, Moscow, 1987) [in
Russian]
\end{thebibliography}
\end{document}